\documentclass[]{pasj02} 
\usepackage[switch,mathlines]{lineno} 
\usepackage{natbib} 
\usepackage{multicol}

\usepackage{multirow}
 \usepackage{caption}
 \usepackage{float}
 \usepackage{url}
 \usepackage{placeins}
 \usepackage{supertabular}

\jyear{2026}
\Received{}
\Accepted{}

\begin{document} 

\title{First and Comprehensive Study of V0757 Pup : \\ $\gamma-$Doradus Pulsator in Detached Eclipsing Binary}

\author{
 B. Danang S. \textsc{Budi},\altaffilmark{1,2}\altemailmark$^{,\dag}$\orcid{0009-0002-4853-9593} \email{danang.budi@grad.nao.ac.jp (Corr. Author)} 
 Bryant \textsc{Randolph},\altaffilmark{3}\orcid{0009-0008-8931-1513}
 Rafa N. \textsc{Akilah},\altaffilmark{1,2}\orcid{0009-0002-0932-8128}
 Evan I. \textsc{Akbar},\altaffilmark{4,5}\orcid{0000-0001-8904-9852}
 Hakim L. \textsc{Malasan},\altaffilmark{4,5,6}\orcid{0000-0001-8549-1811}
 and 
 Puji \textsc{Irawati}\altaffilmark{7}
 }
 
\altaffiltext{1}{Astronomical Science Program, The Graduate University for Advanced Studies, SOKENDAI, 2-21-1 Osawa, Mitaka, Tokyo, 181-8588, Japan}

\altaffiltext{2}{National Astronomical Observatory of Japan, 2-21-1 Osawa, Mitaka, Tokyo, 181-8588, Japan}

\altaffiltext{3}{State Key Laboratory of Dark Matter Physics, Tsung-Dao Lee Institute \& School of Physics and Astronomy, Shanghai Jiao Tong University, Shanghai 201210, People’s Republic of China}

\altaffiltext{4}{Department of Astronomy, Faculty of Mathematics and Natural Sciences, Institut Teknologi Bandung, Bandung 40132, Indonesia}

\altaffiltext{5}{Bosscha Observatory, Institut Teknologi Bandung, Lembang 40391, Indonesia}

\altaffiltext{6}{Department of Atmospheric and Planetary Sciences, Institut Teknologi Sumatera, Lampung Selatan 35365, Indonesia}

\altaffiltext{7}{National Astronomical Research Institute of Thailand, Mae Rim, Chiang Mai 50180, Thailand}



\KeyWords{asteroseismology --- binaries: eclipsing --- stars: fundamental parameters --- stars: individual (V0757 Pup) --- stars: oscillations}  

\maketitle

\begin{abstract}
Pulsating stars in detached eclipsing binary (EA) systems are known for providing many important physical informations which constrain both stellar structure and evolution theories. To date, fewer than 50 $\gamma$ Doradus (GDOR) stars have been found in eclipsing binary systems, making them important targets to study \textit{g}-mode pulsation inside the stars. We present a comprehensive physical and pulsational analysis of V0757 Pup (TIC 6939791), a detached eclipsing binary system observed by the TESS mission (Sectors 7, 34, 61, and 88) and followed up with ground-based spectroscopy from the Thai National Telescope. By combining light curve modeling with radial velocities derived from medium-resolution spectra, we determined the fundamental stellar and atmospheric parameters with high precision. The system consists of an F2V primary ($M_1=1.305\pm0.026M_\odot, R_1=1.643\pm0.020R_\odot$) and a G1V secondary ($M_2=0.934\pm0.030M_\odot, R_2=0.941\pm0.079R_\odot$). We performed a detailed frequency analysis of the residual light curves, identifying two dominant independent pulsation frequencies at $f\sim0.79$ c/d and $0.98$ c/d. These frequencies, along with the derived pulsation constant ($Q\sim0.6$ d), confirm the primary component as a $\gamma$ Doradus pulsator. An analysis of Eclipse Timing Variations (ETV) reveals non detection of third body companions with current available dataset, strengthen by the result of \textit{Gaia} astrometric analysis. Evolutionary modeling indicates the system is $\sim2.2$ Gyr old, with the primary expected to fill its Roche lobe in $\sim1$ Gyr. Additionally, we calculate a distance of $d\approx350$ pc based on orbital and SED modeling, which is in excellent agreement with the \textit{Gaia} DR3 parallax distance.
\end{abstract}


\section{Introduction}

The accurate determination of fundamental stellar parameters—specifically mass, radius, and age—is essential for testing theories of stellar structure and evolution. Detached eclipsing binary (EA) systems have long served as fundamental calibrators for stellar models \citep{1991A&ARv...3...91A,2010A&ARv..18...67T}. By analyzing both the photometric eclipses and radial velocity variations, astronomers can derive absolute masses and radii with better than 1\% precision, independent of stellar atmosphere models. These geometric measurements set strong constraints on physical processes such as convective core overshooting and internal mixing efficiency \citep{2018ApJ...859..100C}.

However, while eclipsing binaries allow precise determination of global stellar properties, they provide limited information about the internal structure. To probe the inner layers and study angular momentum transport, asteroseismology is required. Among main-sequence pulsators, $\gamma$ Doradus (GDOR) stars are particularly valuable. Located near the intersection of the main sequence and the cool edge of the classical instability strip ($T_{\text{eff}} \approx 6900$–$7700$ K) on the Hertzsprung–Russell diagram, these stars pulsate in high-order, low-degree gravity modes (\textit{g}-modes) driven by the convective flux blocking mechanism \citep{1999PASP..111..840K,2000ApJ...542L..57G} with typical pulsational constant $Q>0.23$ d \citep{2002MNRAS.333..251H}. Unlike pressure modes that sample mainly the outer layers, \textit{g}-modes propagate deep into the radiative interior, reaching the boundary of the convective core. Their period spacing patterns provide direct constraints on near-core rotation and the chemical gradient left by core recession—key ingredients in understanding angular momentum evolution \citep{2016A&A...593A.120V,2020MNRAS.497.4363L}.

The discovery of $\gamma$ Doradus pulsators in eclipsing binaries offers a unique opportunity to combine the geometric precision of binary analysis with the internal diagnostics of asteroseismology. In these “seismic binaries,” independent determinations of mass and radius tightly constrain the mean stellar density, which directly scales the pulsation frequency spectrum \citep{2013A&A...556A..56D,2022MNRAS.515.2755S}. This significantly reduces the degeneracy present in seismic modeling of single stars. Moreover, these systems allow investigation of tidal effects on stellar oscillations. Recent studies suggest that tidal interactions can excite pulsation modes or lock them into resonances, influencing both rotational evolution and orbital dynamics \citep{2022MNRAS.511.5860I}.

Despite their strong scientific potential, the number of well-characterized GDOR-EA systems remains small. Although the \textit{Kepler} and TESS missions have identified thousands of pulsating binary candidates \citep{2019A&A...630A.106G,2025Univ...11..302Z}, fewer than 50 systems have been comprehensively analyzed to obtain precise stellar and orbital parameters using high-quality photometry and high-resolution spectroscopy \citep{2018NewA...62...70I,2018MNRAS.480.4693L,2018ApJ...865..115Z,2020MNRAS.491.5980H,2020RMxAA..56..321O,2025MNRAS.538..726Cakirli,2025PASJ..tmp..109L}. This highlights the need for continued follow-up observations and detailed seismic and spectroscopic analyses to fully utilize the capabilities of these important systems.

In this paper, we present a comprehensive analysis of V0757~Puppis (TYC 5405-3070-1, TIC 6939791, $\alpha_{2000}=07^h 33^m 41.^s39$, $\delta_{2000}=-11^\circ 42' 13.''28$, $T_p=+10.46$ mag), which we re-identified serendipitously during the study of another eclipsing binary in its vicinity, as part of the \textbf{BI}naries \textbf{M}inim\textbf{A} monitoring program \citep[BIMA;][]{2015PKAS...30..205H,2024JPhCS2866a2075B}. Although this system is not a new discovery—having been classified as an eclipsing binary in the General Catalogue of Variable Stars \citep[GCVS;][]{1969gcvs.book.....K,2017ARep...61...80S} and listed as an eclipsing binary candidate in the \textit{Gaia} DR3 release \citep{2023A&A...674A...1GaiaDR3}—this work represents the first confirmation of $\gamma$ Doradus-type pulsations in the system, enabled by the high-precision photometry from TESS.

Our goal is to derive the full set of stellar and orbital parameters and to better understand the physical nature of this system through combined analysis of space-based photometry and ground-based spectroscopy. Section~\ref{sec:Observation} describes the data reduction process for the TESS light curve, along with details of the photometric and spectroscopic follow-up observations. The determination of the orbital ephemeris and eclipse timing variations is presented in Section~\ref{sec:ETV}. In Sections~\ref{sec:spectro} and \ref{sec:photo}, we derive the orbital and atmospheric parameters from spectroscopic and photometric modeling, respectively. The residual light curve analysis used to characterize the $\gamma$ Dor pulsations is described in Section~\ref{sec: Residual Analysis}. We briefly discuss the evolutionary status of the system in Section~\ref{sec:evolution}, and summarize our findings in Section~\ref{sec:conclusion}.

\section{Observations}\label{sec:Observation}
\subsection{TESS Photometry}\label{ssec:TESSphot}

TESS observations take consecutive images of the target sky region every 2 seconds, which are co-added to produce several data products, such as Full Frame Images (FFIs), Target Pixel Files (TPFs), and corresponding Light Curves (LCs). Different observation cycles also resulted in various cadence modes; TIC 6939791 was observed during four distinct cycles, all captured in camera 2, as detailed in Table \ref{tab:tess_obs}.

\begin{table}[h]
    \centering
    \caption{\centering TIC 6939791 TESS observation details}
    \begin{tabular}{c c c}
        \hline
        TESS Mission & Year & Cadence Mode (s) \\[2pt]
        \hline
        Sector 07 & 2019 & 1800 \\[2pt]
        Sector 34 & 2021 & 600 \\[2pt]
        Sector 61 & 2023 & 200 \\[2pt]
        Sector 88 & 2025 & 200 \\[2pt]
        \hline
    \end{tabular}
    \label{tab:tess_obs}
\end{table}

\begin{figure*}[h!]
    \centering
    \includegraphics[trim=0mm 0mm 8mm 0mm,clip,width=0.25\linewidth]{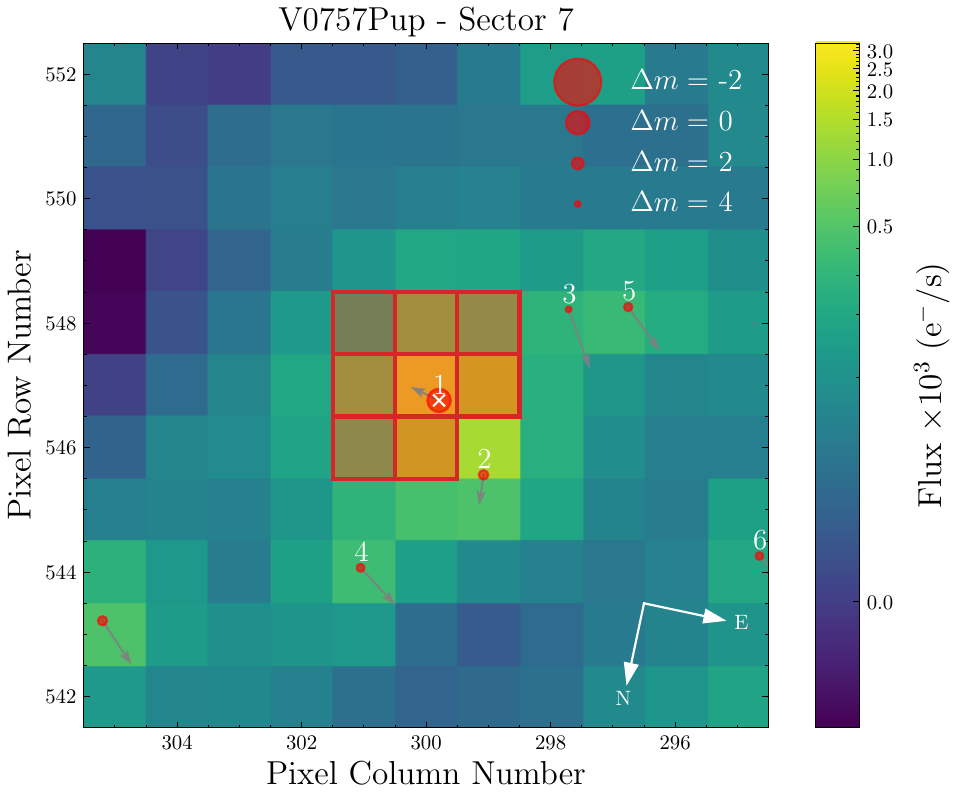}
    \includegraphics[trim=6mm 0mm 8mm 0mm,clip,width=0.24\linewidth]{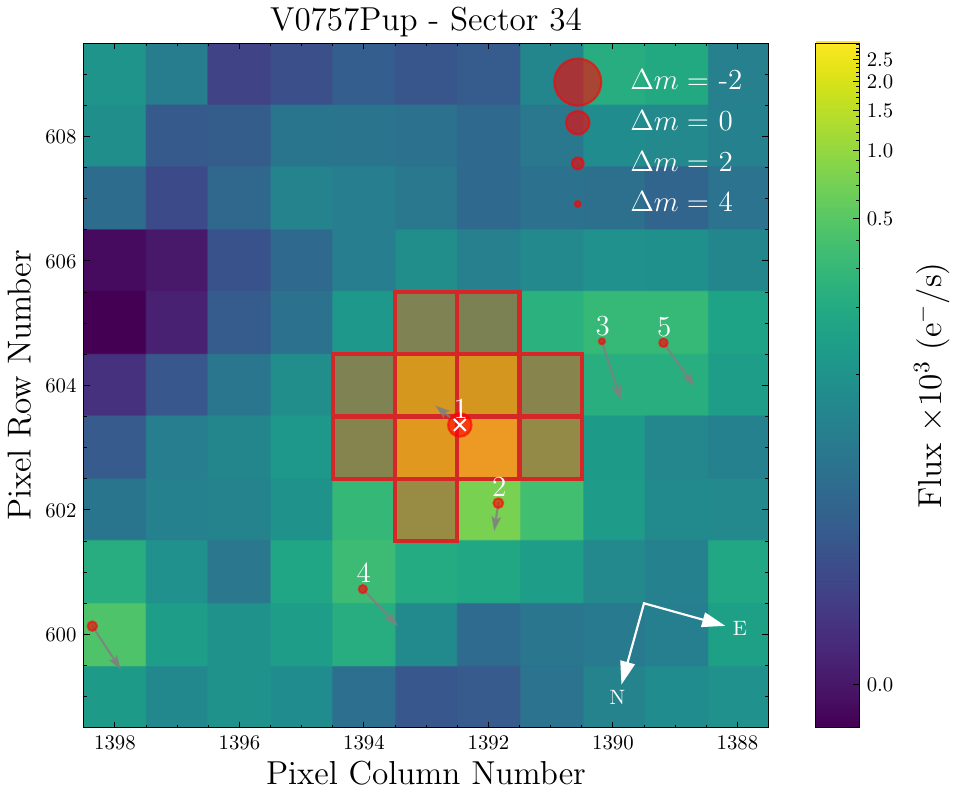}
    \includegraphics[trim=6mm 0mm 8mm 0mm,clip,width=0.24\linewidth]{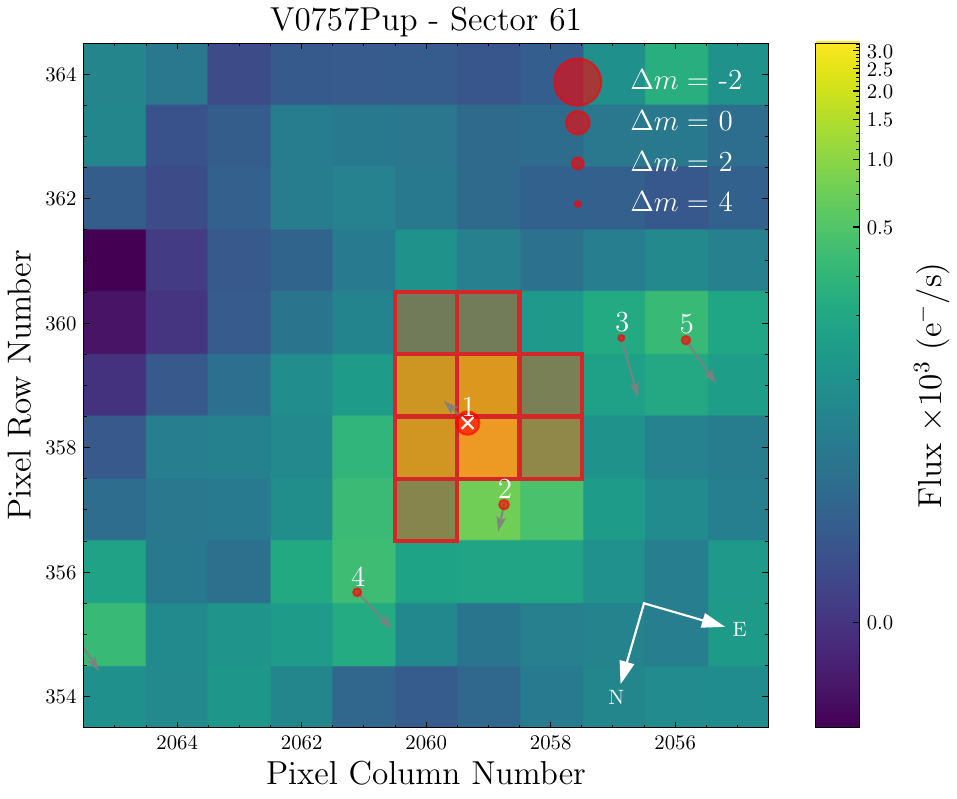}
    \includegraphics[trim=6mm 0mm 0mm 0mm,clip,width=0.25\linewidth]{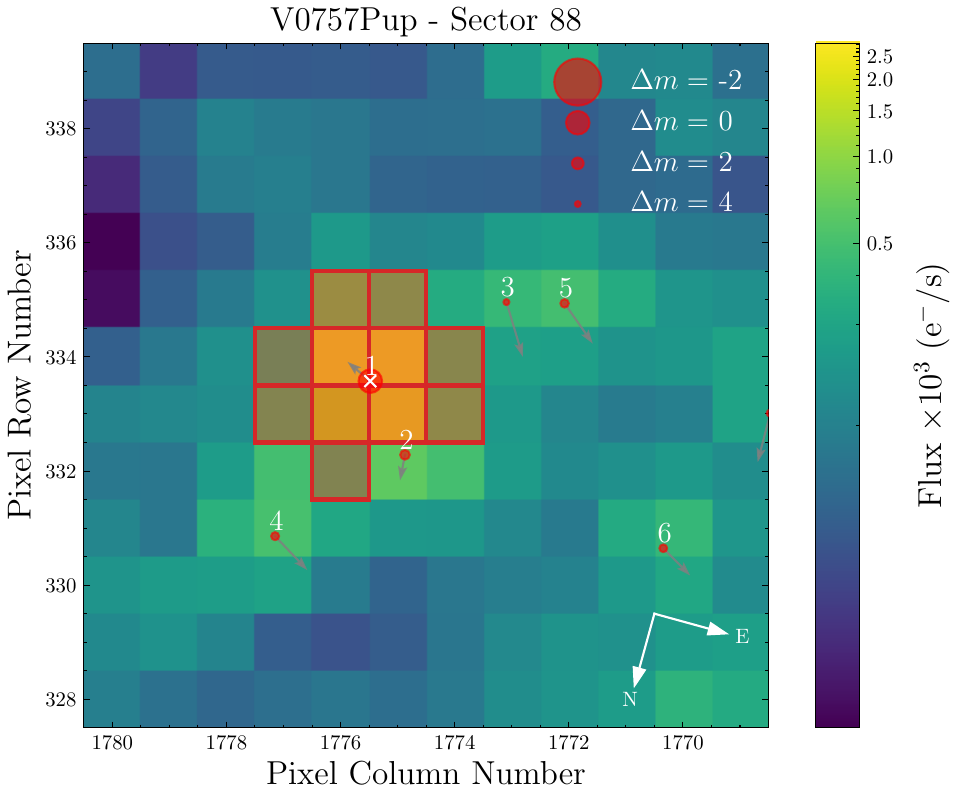}
    \caption{TESS TPFs images of V0757 Pup, observed in Sectors 7, 34, 61, and 88 (displayed from left to right), were generated using \texttt{tpfplotter} \citep{2020aller}. Red dot 1 indicates the position of the target, while the additional red dots represent nearby objects with magnitudes up to 4 mag fainter than the target. The red shaded areas denote the mask applied during flux extraction. Note that object 2, which is close to V0757 Pup ($G=10.46$ mag), corresponds to Gaia DR3 3034136872691540992 ($G=13.01$ mag). Pixels covering this star have already been excluded, although the flux contribution is $\sim 0.28$\%. Gray arrows represent the star's proper motion. \textit{Alt text: TESS images for each sector with indication of pixel used for flux extraction.}}
    \label{fig:TESS_TPFs}
\end{figure*}

\begin{figure*}[h!]
    \centering
    \includegraphics[width=0.8\textwidth]{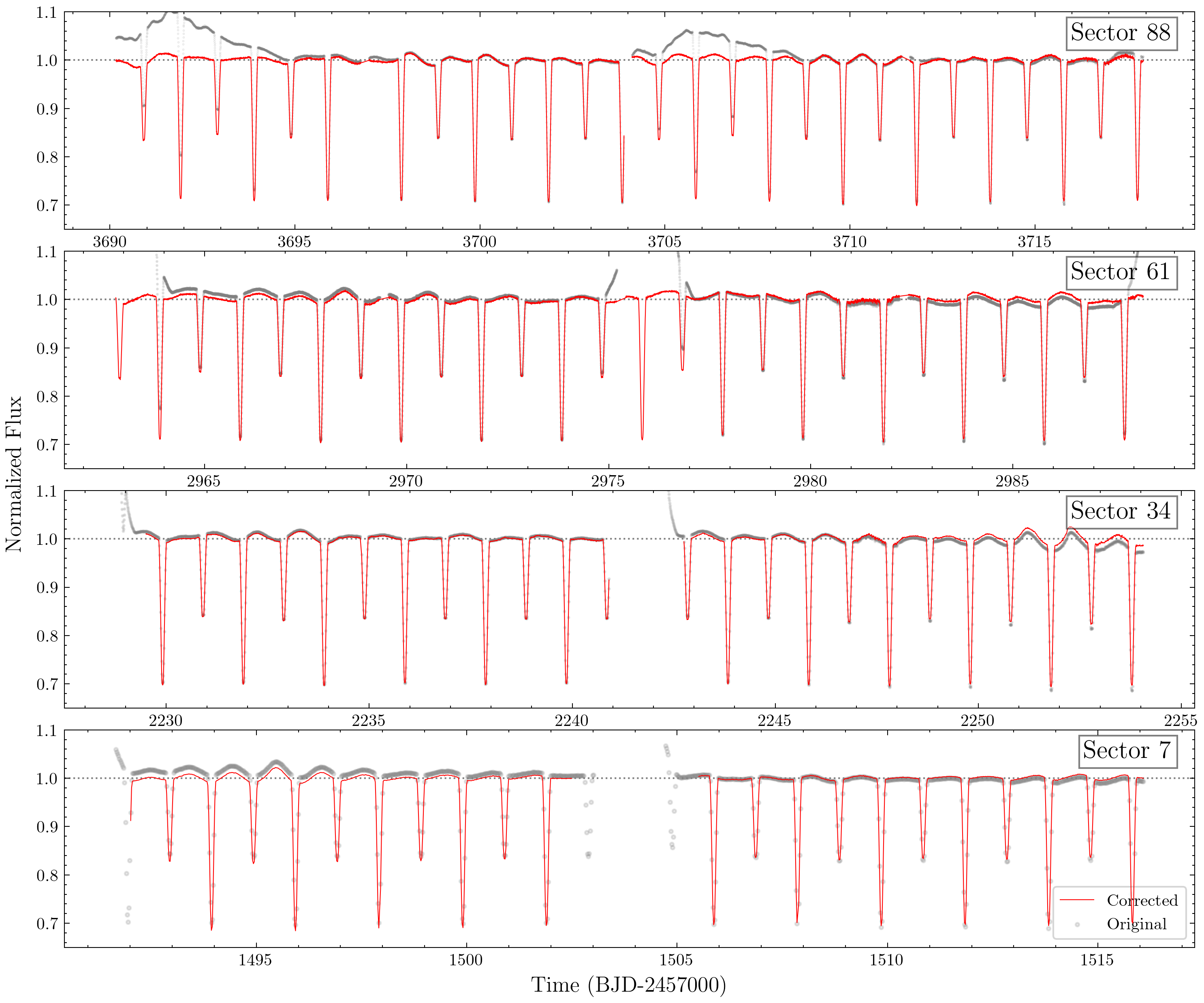}
    \caption{TESS corrected and normalized light curve extracted from aperture as in Figure \ref{fig:TESS_TPFs}.\textit{Alt text: TESS light curve extracted from each sector.}}
    \label{fig:TESS_LCs}
\end{figure*}

We extracted the light curve for each sector from the Mikulski Archive for Space Telescopes \cite[\texttt{MAST}:][]{2019MAST} TESSCut from FFIs with a cut size of $50\times50$, accessible through the package \texttt{LightKurve}\footnote{\url{https://lightkurve.github.io/lightkurve/index.html}} \citep{2018Lightkurve}. We performed simple aperture photometry using this package for all four sectors. The aperture sizes for the source were chosen to avoid pixels where other sources were detected (Figure \ref{fig:TESS_TPFs}). The extracted light curves were then detrended using \texttt{Lightkurve}'s \texttt{RegressionCorrector} class to remove scattered light and spacecraft motion noise from the full $50\times50$ TESS FFIs. The detrended and median-normalized light curves for the four sectors, alongside the aperture used to extract them, are presented in Figure \ref{fig:TESS_LCs}. Sectors 7 and 34 lacked sufficient data points during the observational gap, resulting in suboptimal detrended light curves (especially for Sector 7, which exhibited artificial peaks at the baselines). These artifacts could cause extreme residuals during light curve fitting. Hence, we applied masking to use only data within unaffected observation times for correction and subsequent analysis. As seen in Figure \ref{fig:TESS_LCs}, the baseline experienced a change in brightness (indicated by the presence of consistent periodic brightness variability even after excluding eclipsing events), which we confirmed originated from our target star by altering the masked aperture.

\subsection{TRT-net Photometry}\label{ssec:TRTphot}

Photometric observations from the ground were conducted for V0757 Pup via the Thai Robotic Telescopes network (referred to as \textit{TRT-net}), operated by the National Astronomical Research Institute of Thailand (NARIT), under proposal IDs \texttt{TRTC11A\_009} and \texttt{TRTC11B\_010} (P.I.: BDSB). This robotic telescope network spans the US
(\textit{TRT-SRO}), Chile (\textit{TRT-CTO}), China (\textit{TRT-GAO}), and Australia (\textit{TRT-SRO}). Notably, \textit{TRT-GAO} was not employed in our observations. Comprehensive information about the observation locations, telescope details, and detector specifications can be found on the TRT-net website\footnote{\url{https://trt.narit.or.th/}}, featuring a $10'\times10'$ field of view. The observations were carried out remotely and automatically.
 
The reduction strategy for \textit{TRT-net} photometric observations is as follows: On most observation nights, the \textit{TRT-net} pipeline automatically performs raw image reduction using standard procedures, including bias, dark, and sky flat calibrator images. Additionally, plate solving is conducted to obtain WCS coordinates. In certain cases, manual reduction is required using the calibrators with the \texttt{AstroImageJ} (AIJ) software\footnote{\url{https://www.astro.louisville.edu/software/astroimagej/}} \citep{Collins_2017}. Aperture photometry was performed using AIJ software, and differential photometry was applied to measure relative flux compared to comparison stars (TYC 5405-3142-1; $V\sim11.75$ mag) within the field. However, due to weather conditions and insufficient orbital phase coverage, we opted to use these light curves exclusively for Time of Minima (ToM) measurements.

\subsection{MRES Spectroscopy} 
\label{ssec:mresspectroscopy}
The spectroscopy data used in this work was obtained by the 2.4 m Thai National Telescope (TNT) at the Thai National Observatory (Prop.ID: \texttt{TNTC012\_019}; P.I.: BDSB) during the 2023-2024 observation runs over a total of $\sim$3 nights using the \textbf{M}edium \textbf{R}esolution fiber-fed \textbf{E}chelle \textbf{S}pectrograph (MRES). The MRES spectra have wavelength coverages from 3900-8800\AA\; with spectral resolution $\mathcal{R}\sim16,000-19,000$ depending on the slit configuration \citep{semenko_mkrtichian}. A recent 2025 upgrade on the MRES CCD has expanded its coverage into redder wavelengths $\sim$10,000\AA\;(\textit{private comm.}).

MRES raw spectra were reduced and calibrated with the \texttt{PyYap} pipeline\footnote{\url{https://github.com/ich-heisse-eugene/PyYAP}}, which performed standard 2D image reduction from raw images using bias and flat frames. The wavelength calibration was performed using ThAr lamp emission lines. The wavelength-calibrated spectral orders were combined, and continuum normalization was applied using high-order polynomial fitting.

\section{Eclipse Timing Variation Analysis}\label{sec:ETV}


\subsection{Orbital Linear Ephemeris}\label{ssec:ToM_ephemeris}
We measured the times of minima (ToMs) from the four TESS sectors and the TRT V-band ground-based observations. To obtain consistent results, we used the bisector method \citep{Covino2004} for minima measurements across all datasets. In this approach, we sample the data in the slope region of each minimum and perform a linear extrapolation to determine the midpoint. For the uncertainty measurement, we used the dispersion of the bisector midpoint and applied square-root scaling. All measurements, including the extrapolation, were carried out using the \texttt{SciPy} package \citep{SciPy2020}. Since the relatively high-amplitude $\gamma$-Dor pulsation may significantly distort the shape of the minima, thereby affecting the ToMs, we present here TESS's ToM measurements taken from the \textit{relatively} pulsation-free light curve obtained in Section \ref{sec: Residual Analysis}. For the TRT data, limited light curve coverage prevented us from performing a pulsational analysis. As a result, the data is less reliable, though this is already accounted for by the large derived ToM errors.

From the four TESS sectors and the TRT observation data, we obtained a total of $100$ eclipse minima, consisting of $51$ primary minima and $49$ secondary minima. In addition, we found 12 records of primary minima collected in the O-C gateway\footnote{\url{https://var.astro.cz/en/}}\citep{2006OEJV...23...13P}, which unfortunately lacked ToM uncertainties. Therefore, we excluded them from the analysis and show them only for comparison in the O-C diagram. Together with our data, we obtained ToM data spanning more than 24 years. 

We derived the difference in observed and calculated ToM (O--C) using a preliminary orbital period estimated from the combination of all light curves and the latest primary ToM (in this case TESS Sector 88) through this relation:
\begin{equation}
(O-C)_E = T_E - T_{c,E}=T_E-(T_0+E\cdot P_{orb}),
\end{equation}
where the eclipse cycle number is represented by $E$, and the adopted time of minima and orbital period are symbolized as $T_0$ and $P_{orb}$. 

We detected linear upward trends due to the inaccuracy of the derived periods. From this linear trend, we then updated the linear orbital ephemeris to:
\begin{align}
\text{Min I} = &(60717.7694445948\pm0.0001971868)\;[\text{MJD}]\notag\\&+(1.9891150352\pm0.0000003845)\cdot E,\label{eq:lineph}
\end{align}
which will be used in subsequent analysis.

\begin{figure*}[h!]
\centering
\includegraphics[width=\linewidth]{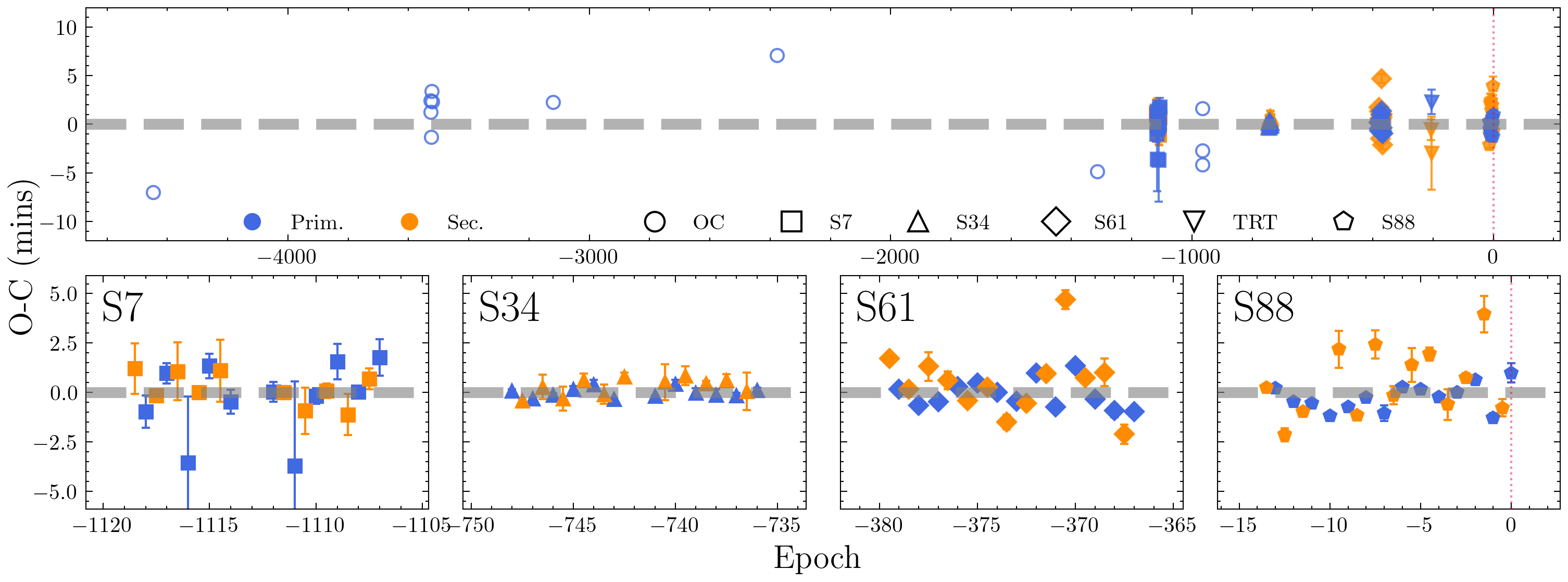}
\caption{Fitting results of linearly corrected O--C diagram shown in upper panel. Blue and orange colors represent primary and secondary eclipse time respectively. Each different markers represent different source of ToM (OC: OC Gateaway, S: TESS). Note that OC Gateaway data did not included in the analysis, hence marked as open circle. Zoom-in version focusing on each TESS sectors are shown in lower panels. \textit{Alt text: OC diagram of V0757 Pup}}
\label{fig:OC_fit}
\end{figure*}

We present the best fit of our ToMs after linear correction in Figure \ref{fig:OC_fit}. All collected minima and derived O–C values used for the analysis are presented in Table \ref{tab:ocdata}. Higher-order trends (parabolic, sinusoidal, etc.) could not be determined using the current data, while small variations ($\lesssim2$ mins) were detected in every TESS sector. We suggest that this might be due to imperfections in our prewhitening process, which may leave some residual pulsation signals that slightly deflect the minima.

\begin{table*}
\begin{center}
\caption{Times of minima and O--C measurements.}
\label{tab:ocdata}
\small 
\setlength{\tabcolsep}{2.5pt}
\begin{tabular}{cccccccc}
\hline
Time & Epoch & O--C & Error & I/II & Phot. Band & Source & Observer \\[2pt]
(BJD$-2.4\times10^6$) & & (min) & (min) & & & &  \\[2pt]
\hline
51870.1809 & -4448   & -7.0119 & --     & I  & V & OC G (N/A)         & Pojmanski G \\[2pt]
53702.1624 & -3527   & 2.4458  & --     & I  & V & OC G (VSOLJ 0044) & Nakajima Kazuhi \\[2pt]
53704.1507 & -3526   & 1.2384  & --     & I  & V & OC G (VSOLJ 0044) & Nakajima Kazuhi \\[2pt]
53708.1271 & -3524   & -1.3228 & --     & I  & V & OC G (VSOLJ 0044) & Nakajima Kazuhi \\[2pt]
53710.1195 & -3523   & 3.3713  & --     & I  & V & OC G (VSOLJ 0044) & Nakajima Kazuhi \\[2pt]
53712.1079 & -3522   & 2.3039  & --     & I  & V & OC G (VSOLJ 0044) & Nakajima Kazuhi \\[2pt]
54509.7430 & -3121   & 2.2511  & --     & I  & V & OC G (IBVS 6029)  & Diethelm Roger \\[2pt]
55987.6588 & -2378   & 7.1033  & --     & I  & V & OC G (IBVS 6029)  & Diethelm Roger \\[2pt]
58102.0798 & -1315   & -4.8495 & --     & I  & V & OC G (N/A)         & Kazarovets E \\[2pt]
58492.9451 & -1118.5 & 1.2074  & 1.2773 & II & T & Sec 7             & TESS \\[2pt]
58493.9382 & -1118   & -0.9768 & 0.8202 & I  & T & Sec 7             & TESS \\[2pt]
$\vdots$ & $\vdots$ & $\vdots$ & $\vdots$ & $\vdots$ & $\vdots$ & $\vdots$ & $\vdots$ \\[2pt]
\hline
\end{tabular}
\end{center}
\begin{tabnote}
I: Primary Eclipses, II: Secondary Eclipses, OC G: Taken from OC Gateaway. The full, untruncated version of this table is available online (see Supplementary Data).
\end{tabnote}
\end{table*}

\subsection{Is there a third companion in this system?}\label{ssec:discussion}
Despite the non-detection of higher-order ToM deviations in the previous section, there remains a possibility of a low-mass third-body companion that exhibits longer sinusoidal variation. The limited O–C coverage during the pre-TESS era and the numerous gaps between campaigns make such a signal challenging to detect \citep{Deeg2008,2021RMxAA..57..419P}.

We further investigated this scenario through astrometric data. A search within a 1$''$ radius around V0757 Pup using \textit{Gaia} DR3 revealed no nearby sources (within the \textit{Gaia} magnitude limit). We also tested for proper motion anomalies (PMa) using the \textit{Gaia} DR2 \citep{2018A&A...616A...1G} and DR3 \citep{2023A&A...674A...1GaiaDR3} catalogues, following the methods of \citet{Brandt2021} and \citet{Kervella2019}, through Eq.\ref{eq:PMaSignificance}, where $\rho_i$ denotes the correlation between $\mu_{\alpha*}$ and $\mu_\delta$ for each release. Using the values listed in Table~\ref{tab:pmGAIA}, we obtain a PMa significance of $S_\text{PMa} = 2.56\sigma$—suggestive, but below the standard $3\sigma$ detection threshold.

\begin{table}[h!]
\centering
\caption{Proper motion value of V0757 Pup with GaiaID 3034136838331806720 as reported by Gaia mission}
\label{tab:pmGAIA}
\begin{tabular}{lcc}
\hline
 & $\mu_{\alpha_*}$ (mas/yr) & $\mu_\delta$ (mas/yr) \\[2pt]
\hline
DR2 & $-8.5356 \pm 0.0562$ & $-4.1785 \pm 0.0492$ \\[2pt]
DR3 & $-8.4499 \pm 0.0125$ & $-4.0735 \pm 0.0123$ \\[2pt]
\hline
\end{tabular}
\end{table}

Since the source is already a binary system, an additional companion would need to induce reflex motion comparable to or larger than that of the binary itself to be clearly detectable. Thus, the PMa measurement only weakly supports the possibility of a third object, which would likely need to be either a low-mass star or a massive S-type planet. With current astrometric precision, however, the evidence remains inconclusive.

\begin{equation}
\label{eq:PMaSignificance}
\begin{aligned}
\Delta \boldsymbol{\mu} &= 
\begin{bmatrix}
\Delta \mu_{\alpha_{*}} \\
\Delta \mu_\delta
\end{bmatrix} 
= 
\begin{bmatrix}
\mu_{\alpha_{*},\mathrm{DR3}} - \mu_{\alpha_{*},\mathrm{DR2}} \\
\mu_{\delta,\mathrm{DR3}} - \mu_{\delta,\mathrm{DR2}}
\end{bmatrix}, \\[2mm]
\mathbf{C}_\mathrm{tot} &= \mathbf{C}_\mathrm{DR2} + \mathbf{C}_\mathrm{DR3}, \\[2mm]
\mathbf{C}_\mathrm{DRi} &=
\begin{bmatrix}
\sigma_{\mu_\alpha{_{*}},i}^2 & \rho_i \, \sigma_{\mu_\alpha{_{*}},i} \, \sigma_{\mu_\delta,i} \\
\rho_i \, \sigma_{\mu_\alpha{_{*}},i} \, \sigma_{\mu_\delta,i} & \sigma_{\mu_\delta,i}^2
\end{bmatrix}, \\[1mm]
\chi^2 &= \Delta \boldsymbol{\mu}^\mathrm{T} \, \mathbf{C}_\mathrm{tot}^{-1} \, \Delta \boldsymbol{\mu}, \quad
S_\mathrm{PMa} = \sqrt{\chi^2}.
\end{aligned}
\end{equation}

\section{Spectroscopic Analysis}\label{sec:spectro}
\subsection{Radial Velocity Measurement}\label{ssec:RVmeasurement}

To derive radial velocities, we performed Least Squares Deconvolution \citep[LSD,][]{1997MNRAS_donati} analysis using the code described in \citet{2010AA_kochukhov}. A line mask containing several atomic lines with their expected line strengths for F-class solar-metallicity spectra was prepared for these calculations, avoiding strong wing regions such as Balmer lines and regions with telluric contamination. This line mask was extracted from the Vienna Atomic Line Database \citep[\texttt{VALD},][] {1995A&ASvald1,2000BaltA.vald2,2015PhySvald3}.

\begin{figure}[h!]
    \centering
    \includegraphics[trim=4mm 10mm 4mm 12mm,clip,width=\linewidth]{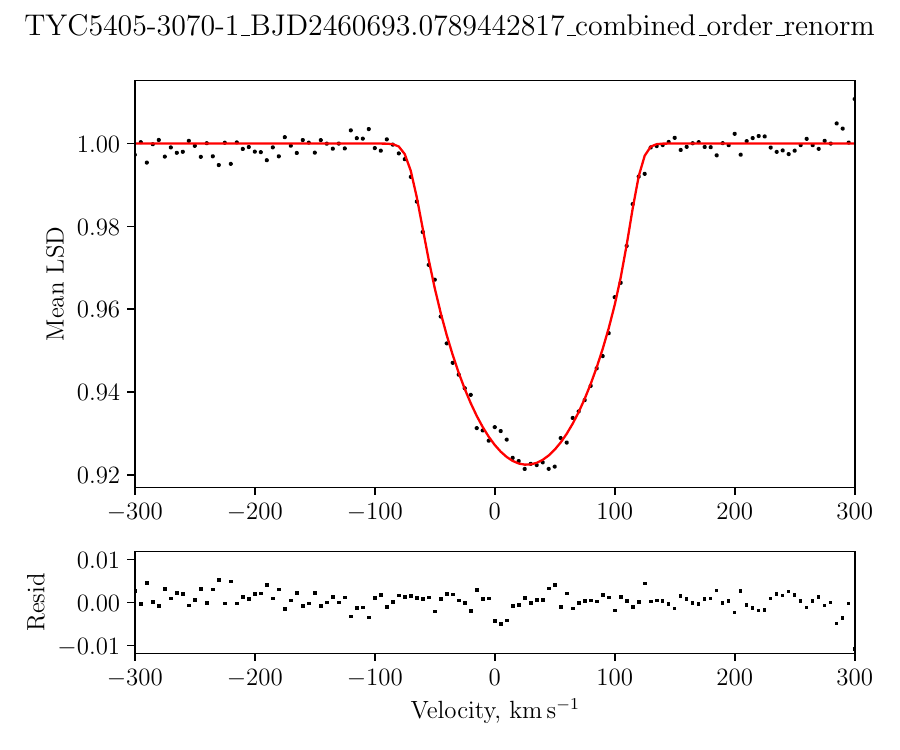}
    \vspace{-0.5mm}
    \includegraphics[trim=4mm 3mm 4mm 12mm,clip,width=\linewidth]{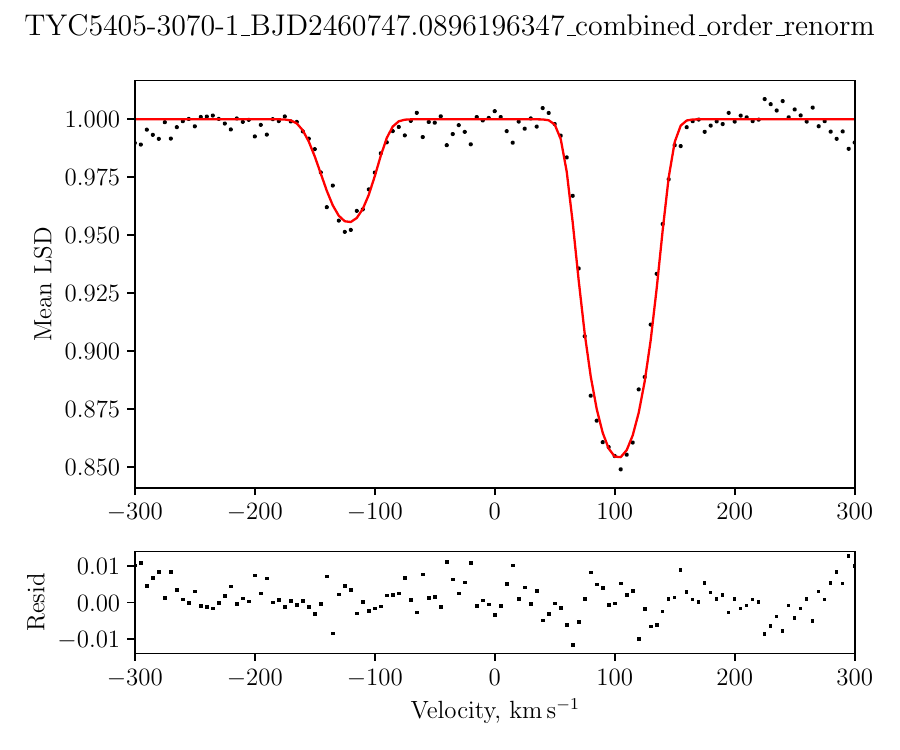}
    \vspace{+3mm}
    \caption{The mean LSD profiles (black dots) obtained from MRES spectra for V0757 Pup at BJD = 2460693.0789442817 and 2460747.0896196347 correspond to the upper and lower panels, respectively. The red solid line represents the fitted LSD profile used to determine $v \sin i$ and radial velocity. Notice that the upper panel, which corresponds to the orbital phase $\phi\sim0.58$, is heavily blended between the primary and secondary components; hence, we could not determine the velocity of each component, compared to the lower panel at $\phi\sim0.74$. \textit{Alt text: LSD profile fitting results for two epoch of V0757 Pup observation.}}
    \label{fig:lsdfit}
    \vspace{-0.5cm}
\end{figure}

The resulting LSD profiles were then fitted with a rotational broadening function \citep{2005oasp.book_gray}, where the full width at half maximum (FWHM) depends on the stellar projected rotational velocity ($v \sin i$), and the profile center depends on the stellar radial velocity. When the radial velocities of each component are sufficiently well-separated, we observe two distinct absorption features within the LSD profile. These features can be fitted separately to yield the projected rotational ($v\sin i$) and radial velocities for each component. In addition, the LSD profile depth ratio between the primary and secondary components can be used to estimate light contribution. Example fitting results are shown in Figure \ref{fig:lsdfit}, and the resulting RV, $v\sin i$, light fraction for each component, along with the signal-to-noise ratio (S/N) of a 10Å line-free region around 600nm, are provided in Table \ref{tab:RV} in Appendix \ref{appendix:RV_table_output}. Notably, one spectrum was heavily blended; this RV data point was excluded from subsequent analysis (see Fig. \ref{fig:lsdfit}, upper panel).

\subsection{Spectral Disentangling}\label{ssec:spec_dist}

To model the light curve, we first need to determine the primary effective temperature. Initially, from color indices $B-V=0.34\pm0.08$ \citep{2000A&A_tycho2}, we obtain an estimate of $T_\text{eff, prim}\approx6800$ K \citep{2005oasp.book_gray}. We performed preliminary modeling on light and radial velocity, yielding an estimate for the light contribution of each component of $L_s/L_p\sim0.211$, consistent with the LSD profile fitting result of $\text{lf}_s/\text{lf}_p\sim0.22-0.30$ (see Table \ref{tab:RV}). We then used the light contribution estimated from LSD profile fitting to improve the primary temperature estimation through spectral disentangling.

Spectral disentangling (SD) is a technique used to separate individual spectra for each stellar component from composite spectra \citep{1994simon,1995Hadrava}, under the assumption that there is no intrinsic variability within their line profiles. Good phase coverage and a high S/N ratio are required for this process. In this analysis, we used a \texttt{PYTHON}-based wrapper, \texttt{fd3\_initiator}\footnote{\url{https://github.com/ayushmoharana/fd3_initiator}} \citep{2023Moharana}, for the disentangling code \texttt{FDBINARY} \citep{2004ASPC_Ilijic}. We provided the orbital ephemeris ($T_\text{periastron}$ and $P_\text{orb}$) along with the RV semi-amplitudes ($K_1$ and $K_2$) from the RV modeling solution as inputs, fixing these values due to the limited phase coverage of our data and assuming a circular orbit. Additionally, one spectrum was excluded due to its low S/N. The spectral range was set for two regions of interest: around H$\beta$ and the Mg triplets with a bandwidth of $\sim125$\AA. The disentangling process was performed independently for both regions. The resulting RVs from both regions were then compared with those obtained from LSD profile fitting, showing consistent values with deviations of $<10$ km/s and $<15$ km/s for the primary and secondary components, respectively. These SD-derived RVs were not used for the RV analysis. The disentangled spectra were then continuum-corrected using the \texttt{SUPPNet}\footnote{\url{https://rozanskit.com/suppnet/}} package \citep{2022AA_suppnet}.

We estimated the spectral flux uncertainties $(\sigma_\text{Flx})$ through the relation:
\begin{align}
\sigma_\text{Flx}=\sqrt{\sigma_\text{SD}^2+\sigma_\text{cont}^2}\;\;,
\end{align}
where $\sigma_\text{SD}$ represents flux uncertainties resulting from the spectral disentangling process and $\sigma_\text{cont}$ are continuum uncertainties estimated from \texttt{SUPPNet}.

\subsection{Determination of Atmospheric Parameters and Chemical Abundances of the Primary Component}\label{ssec:stellarparam}

To achieve higher precision in temperature determination, we conducted spectral modeling on both component-normalized, disentangled spectra. We focused on the $T_\text{eff}$–$\log g$ sensitive Balmer lines, which are well-suited for AF-class stars. The disentangled spectra cover both H$\alpha$ and H$\beta$ lines, with visual inspection revealing a notably broad wing in the primary spectrum.

However, due to the lower S/N ratio and continuum correction issues in the H$\alpha$ line of the primary spectrum, we excluded this line from the analysis. The secondary component spectra also suffered from low S/N and continuum placement uncertainties in both Balmer lines, which prevented accurate recovery of their true profiles. The absence of broad wings in the Balmer lines of the secondary component further indicates that this star is not highly sensitive to changes in $T_\text{eff}$ and $\log g$.

Our preliminary analysis of the secondary spectrum indicates significantly lower $T_\text{eff}$ and $\log g$ values compared to the photometric results. Therefore, we refrained from performing further spectral analysis on the secondary component.

First, we derived the $T_\text{eff}$, $\log g$, and rough estimates of metallicity for the primary spectra by performing spectral synthetic fitting over a $\sim120$ \AA\; wavelength region around H$\beta$. We used the 1D \texttt{MARCS} model atmosphere \citep{2008AA...MARCS} and the radiative transfer code \texttt{TurbospectrumNLTE} \citep{2012ascl.TS,2023AA...gerber_TSNLTE} via the \texttt{PYTHON}-based wrapper \texttt{TSFitPy}\footnote{\url{https://github.com/TSFitPy-developers/TSFitPy}} \citep{2023MNRAS_TSFitypy}. To remove the effect of absorption lines on the fitting process of the Balmer wing, we defined several "continuum line-free" regions and synthesized the spectra using only the H line list. Since the core regions of Balmer lines are significantly affected by non-LTE effects and saturation, we excluded a 2 \AA\; region around the line cores during the analysis. Although the wings of Balmer lines are less affected by non-LTE effects, we applied non-LTE corrections \citep{2008Amashonkina_HNLTE} to improve accuracy. The synthetic spectra were then convolved to match the MRES spectral resolution and the fixed $v \sin i$ determined via LSD profile fitting.

\begin{figure*}[ht!]
    \centering
    \includegraphics[width=\linewidth]{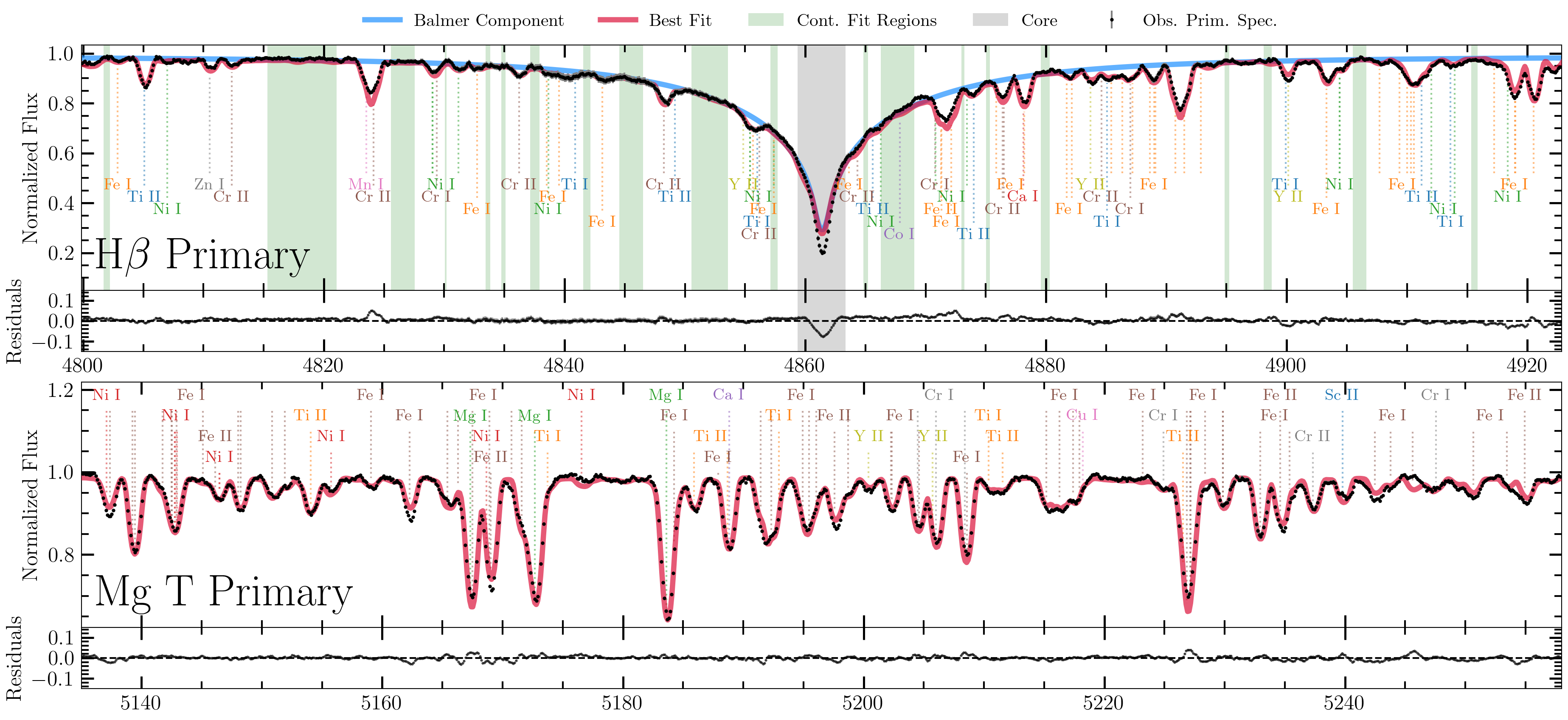}
    \caption{Upper and lower panels display the disentangled spectra of regions around H$\beta$ and Mg triplets for primary component. Data point represent spectra and its uncertainty. The solid red lines is the best fit spectra adopting all derived atmospheric parameters and abundances listed on Table \ref{tab:best_fit_spectral}. The solid blue line are the best fit H$\beta$ component only. We indicated the regions for pseudo-continuum fitting of Balmer line as green shades regions and grey shade region is the H$\beta$ core that we excluded in the analysis. Lower panels in each subplot show the residual. \textit{Alt text: balmer line and Mg triplet regions of primary component of V0757 Pup with best fitting from spectral analysis.}}
    \label{fig:fittedspectra}
    \vspace{-0.4cm}
\end{figure*}

For parameter optimization, we used Markov Chain Monte Carlo (MCMC) sampling with the \texttt{emcee} sampler \citep{2013PASP..125..306F}, maximizing the total log-likelihood. We assumed flat priors for temperature ($\mathcal{U}[4000,8000]$ K) and surface gravity ($\mathcal{U}[1,5]$), and a Gaussian prior for metallicity $(\mathcal{G}[-0.5,1])$, based on estimates from \texttt{GSP-Phot Aeneas} in \textit{Gaia} DR3 \citep{2011MNRASbailer}. The simulation was run until convergence with 75 walkers over 600 iterations, discarding the first 250 iterations as burn-in. The best-fitted Balmer component is shown as a solid blue line in the upper panel of Figure \ref{fig:fittedspectra}, and posterior distributions are shown in Figure \ref{fig:cornerbalmer} (upper right corner plots) in Appendix \ref{appendix:corner_plot}.

To better constrain the Iron abundance (metallicity) and other elemental abundances, we employed a simultaneous fitting for Fe and 7 other elements based on several atomic lines detected in a 125Å window around the Magnesium triplet lines (5135-5260Å). The low spectral resolution, large rotational velocity, and near-solar metallicity of this star widen the spectral lines and blend nearby atomic features, making individual line-by-line analysis highly challenging. Therefore, as a first approximation, modeling the entire region at once while varying individual elemental abundances yields better results. The recurrence of certain atomic lines in this region helps prevent degeneracies when modeling blended lines. We found it highly challenging to set $T_\text{eff}$ and $\log g$ as free parameters while simultaneously fitting 8 elemental abundances and other parameters (e.g., RV, continuum placement, microturbulence velocity, and $v \sin i$). Hence, we fixed $T_\text{eff}=7033$ K and $\log g=4.00$, based on the prior H$\beta$ analysis.

For optimizing all 11 parameters together, we again employed MCMC sampling with the \texttt{emcee} sampler. We assumed Gaussian priors for metallicity $(\mathcal{G}[-0.5,0.75]\text{ dex})$, $v \sin i$ $(\mathcal{G}[40,10]\text{ km s}^{-1})$ based on the LSD analysis, RV $(\mathcal{G}[0,10] \text{ km s}^{-1})$, and continuum placement $(\mathcal{G}[1,0.05])$, while setting a flat prior of [X/Fe]\footnote{$\text{[A/B]} = \log(N_\text{A}/N_\text{B}) - \log(N_\text{A}/N_\text{B})_\odot$, and $A(\text{X}) = \log\epsilon_\text{X} = \log(N_\text{X}/N_\text{H}) + 12$, where A, B, and X are elements.} ($\mathcal{U}[-0.5,+2.0]$ dex) for other elements. We adopted line lists from the Gaia-ESO-Survey \citep[\texttt{GES} :][]{2021A&A...645A.106H}, assuming LTE conditions. Solar abundances were adopted from \cite{2022A&A...661A.140M}.

The simulation was run until convergence with 75 walkers over 2000 iterations, discarding the first 1500 iterations as burn-in. The best-fitted spectra are shown as a solid red line in the lower panel of Figure \ref{fig:fittedspectra}. We also synthesized the full spectra around the H$\beta$ region using the adopted parameters, shown as a solid red line in the upper panel. Posterior distributions are plotted in Figure \ref{fig:cornerbalmer} (lower left) in Appendix \ref{appendix:corner_plot}. The derived atmospheric parameters and chemical abundances for the primary component, along with their propagated uncertainties, are listed in Table \ref{tab:best_fit_spectral}. We acknowledge that the abundance uncertainties are derived purely from flux uncertainty ($\sigma_\text{Flx}$) and thus may underestimate true errors. A more robust uncertainty derivation would require line-by-line analysis and incorporate $T_\text{eff}$ and $\log g$ uncertainties, which is beyond the scope of this study.

\begin{table}[h!]
\centering
\caption{Best fit atmospheric parameters and chemical abundances for Primary Component}
\label{tab:best_fit_spectral}
\begin{tabular}{lc}
\hline
\multicolumn{1}{c}{\textbf{Parameters [unit]}} & \textbf{Value} \\[2pt]\hline
$T_\text{eff}$ [K] & $7033^{+54}_{-80}$ \\[2pt]
$\log g$ [cgs] & $3.988^{+0.105}_{-0.104}$ \\[2pt]
$v\sin i$ [km s$^{-1}$] & $41.201^{+0.031}_{-0.032}$ \\[2pt]
$v_\text{mic}$ [km s$^{-1}$] & $3.369\pm0.012$ \\[2pt]
 & \\[2pt]
 $\text{[Fe/H]}_\text{LTE}$ [dex] & $-0.625\pm0.002$ \\[2pt]
$\text{[Ni/Fe]}_\text{LTE}$ [dex] & $+0.005\pm0.005$ \\[2pt]
$\text{[Ti/Fe]}_\text{LTE}$ [dex] & $+0.258\pm0.004$ \\[2pt]
$\text{[Mg/Fe]}_\text{LTE}$ [dex] & $+0.479\pm0.002$ \\[2pt]
$\text{[Ca/Fe]}_\text{LTE}$ [dex] & $+0.385\pm0.010$ \\[2pt]
$\text{[Cr/Fe]}_\text{LTE}$ [dex] & $+0.187\pm0.007$ \\[2pt]
$\text{[Y/Fe]}_\text{LTE}$ [dex] & $-0.131\pm0.026$ \\[2pt]
$\text{[Sc/Fe]}_\text{LTE}$ [dex] & $+0.619\pm0.019$ \\[2pt]\hline
\end{tabular}
\end{table}

In general, the synthetic spectra for the Mg T region reproduce the observed spectra well, except for minor differences around 5240-5248Å, possibly due to unidentified atomic lines or non-LTE effects. For the H$\beta$ region, the synthetic spectra and Balmer wings also reproduce the observations well, with the exception of some blended lines, such as the 4823Å Mn I and Cr II feature. We note that Mn was not included in our Mg T region analysis, and we assumed solar abundance ($\text{[Mn/Fe]}=0$) for all non-derived elements. Lowering the Mn abundance might improve the fit. We do not plan further detailed abundance analysis for other lines in this work. Future studies utilizing higher spectral resolution than MRES will be necessary for weak line abundance analysis (e.g., Lithium).

Our spectroscopic analysis classifies the system's primary as an F2V star. We compared this result with previous studies. The TESS Input Catalogue v8.2 \citep{2022yCat_TESSv82} provided estimates of $T_\text{eff} = 6737 \pm 136$ K and $\log g = 4.02 \pm 0.09$, based on \textit{Gaia} DR2 \citep{2018A&A...616A...1G}. The updated \textit{Gaia} catalogue reports $T_\text{eff} = 6838.7^{+11.9}_{-15.7}$ K and $\log g = 3.97 \pm 0.01$, derived from \texttt{GSP-Phot Aeneas} \citep{2011MNRASbailer}. Overall, our measured effective temperature is $\sim200-300$ K higher, with a similar surface gravity, compared to prior work. Several factors support the reliability of our measurements, notably the use of non-LTE spectral models for the Balmer lines, compared to the single-star assumptions in earlier catalogs.

We also note that the $v_p\sin i$ resulting from the spectral analysis is consistent with, and provides tighter constraints than, the value obtained from averaging the $v_p\sin i$ obtained from LSD profile analysis (see Table \ref{tab:RV}). The microturbulence velocity for the primary component is consistent with typical main-sequence F-class stars \citep{2025A&A...701A.297M}.

Henceforth, the primary effective temperature will be treated as an invariant parameter in all subsequent analyses. We observed an inconsistency regarding the Fe abundance compared to the one resulted from isochrone analysis (discussed further in Section\ref{sec:evolution}). Future works are necessary to refine the chemical abundance analysis using higher-resolution spectra.

\subsection{Interstellar Extinction Estimation}\label{ssec:reddening}
\begin{figure}[h!]
    \centering
\includegraphics[width=\linewidth]{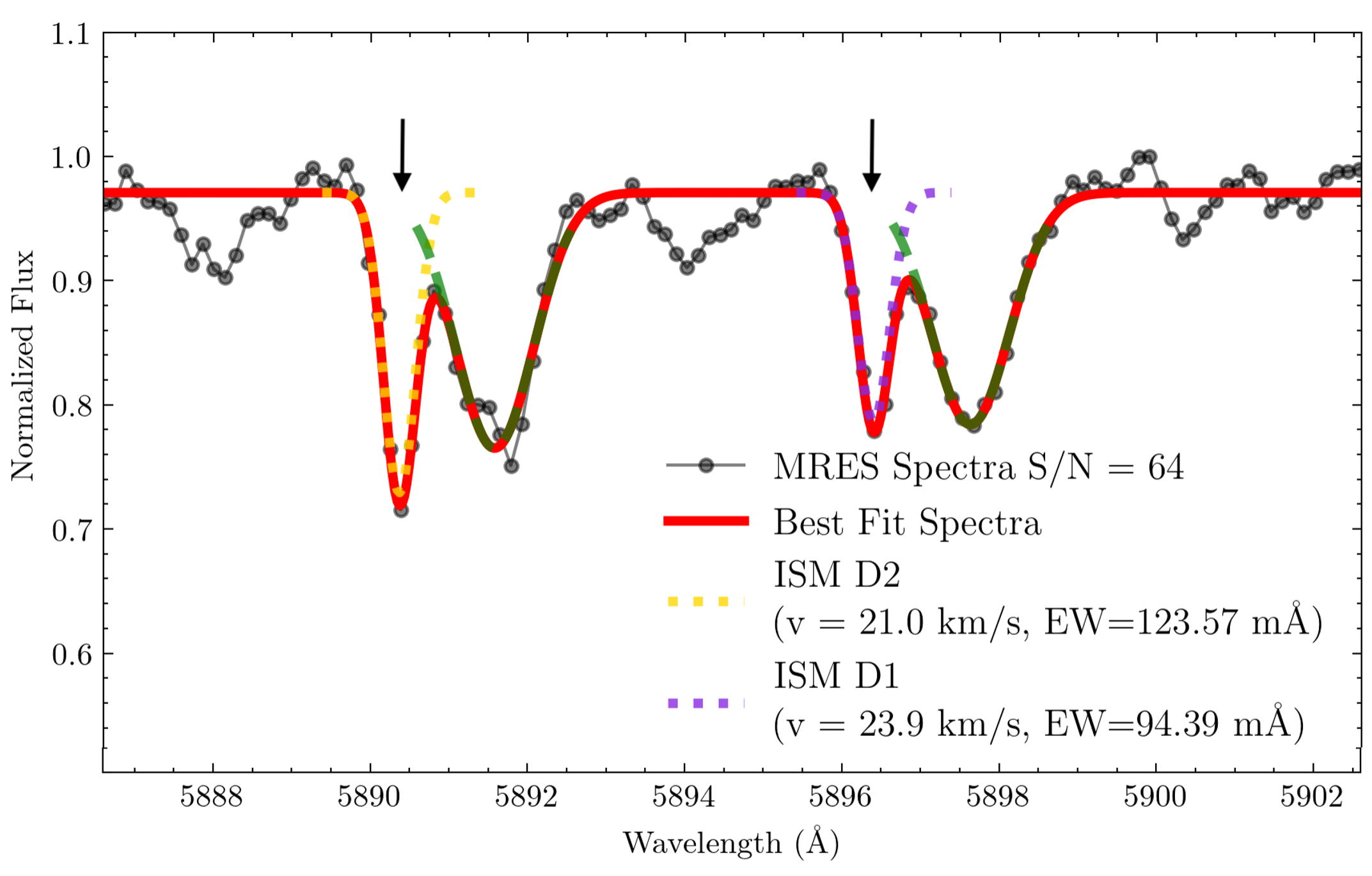}
    \caption{Example result of multi-gaussian fit around $\text{Na I D}_1$ \& D$_2$ of MRES spectra (black dots). The narrower line marked by arrows are interstellar lines while broader Na lines are stellar component. \textit{Alt text: Example spectra of V0757 Pup around sodium doublet lines.}}
    \label{fig:NaD_fit}
\end{figure}

An additional advantage of sufficiently high spectral resolution is its ability to detect and separate the sodium doublet lines (Na I D $\lambda\lambda5890;5896$ \AA) associated with stellar and interstellar components. In this study, we utilized high-S/N MRES spectra to estimate interstellar extinction (reddening) along the line of sight towards V0757 Pup. From a total of 16 spectra, we reliably measured the equivalent widths (EWs) of interstellar $\text{Na I D}_1$ and D$_2$ lines in 13 spectra covering various orbital phases. This was achieved through multi-Gaussian fitting, which effectively distinguished them from stellar absorption lines as illustrated in Figure \ref{fig:NaD_fit}. The average EWs and their standard deviations were calculated as $84.53 \pm 19.72$ m\AA; and $129.42 \pm 14.54$ m\AA; for the D$_1$ and D$_2$ absorption lines, respectively.

To estimate the reddening parameter $E(B-V)$, we applied Equations (7) -- (9) from \citet{2012MNRAS_NaD}, which established an empirical relationship based on over 100 high-resolution spectra from the Keck telescope and nearly a million low-resolution spectra from the Sloan Digital Sky Survey (SDSS). A Monte Carlo simulation was conducted using input parameters and associated uncertainties, drawing median values along with the 16th and 84th percentiles from $10^6$ samples. The resulting reddening values were $E(B-V) = 0.028^{+0.014}_{-0.009}$, $0.023^{+0.010}_{-0.007}$, and $0.025^{+0.005}_{-0.004}$ mag, derived from the EWs of $\text{Na I D}_1$, D$_2$, and combined D$_1$+D$_2$ lines, respectively. The weighted average adopted for subsequent calculation is $\langle E(B-V)_\text{spec}\rangle=0.02563^{+0.00415}_{-0.00379}$ mag.

We compared these results with those obtained via other methods. First, using the 2D Galactic Dust and Reddening Map\footnote{\url{https://irsa.ipac.caltech.edu/applications/DUST/}} from \citet{2011ApJ_SF}, we obtained a significantly higher value of $E(B-V) = 0.291 \pm 0.007$ mag. However, considering that this system is relatively nearby, with a parallax distance of $\sim$352 pc \citep{2023A&A...674A...1GaiaDR3}, we applied a location correction to account only for the dust in the foreground, following \citet{2000AJ....120.2065Bonifacio}. This adjustment yielded an estimated $E(B-V) \sim 0.05$ mag, aligning closer with our spectroscopic estimation and the value of $E(B-V) = 0.040 \pm 0.011$ from the TESS Input Catalogue v8.2 \citep{2022yCat_TESSv82}. Additionally, we compared our results with data from the 3D Dust Mapping project\footnote{\url{http://argonaut.skymaps.info}} \citep{2019ApJ...887...93Green2019}, obtaining $E(B-V) = 0.981 E(g-r) \approx 0.02 \pm 0.02$ mag. It is important to note that this value is extrapolated due to insufficient data within the relevant distance range. Our independent reddening measurement based on Na I doublet lines ultimately confirms the results of previous studies.

\section{Photometric Analysis}\label{sec:photo}
\subsection{Simultaneous Radial Velocity and TESS Light Curve Modeling}\label{ssec:rv_photo_fitting}

After obtaining the RV data for each component from the LSD analysis (Section \ref{ssec:RVmeasurement}) and the cleaned TESS light curves (Section \ref{ssec:TESSphot}), simultaneous RV and light curve modeling were performed using the \texttt{PyWD2015}\footnote{\url{https://github.com/Varnani/pywd2015-qt5}} program \citep{pywd_GUI}. This software is based on the Wilson-Devinney \citep[W-D:][]{pywd_1, pywd_2, pywd_3, pywd_4, pywd_5} algorithm and offers a user-friendly, \texttt{PYTHON}-based graphical interface. As mentioned in Section \ref{ssec:ToM_ephemeris}, we initially performed rough RV and LC modeling to get the best $P_\text{orb}$ across all RVs and LCs. Then, using the preliminary $P_\text{orb}$, $T_0$, and ToM data, we refined $P_\text{orb}$ and $T_0$, yielding the Linear Ephemeris of Equation \ref{eq:lineph}.

After adopting the new ephemeris and setting them as fixed parameters, we began by modeling the RV curves to constrain several RV-sensitive parameters (e.g., semi-major axis $a$, systemic velocity $\gamma_\text{sys}$, and mass-ratio $q\equiv M_s/M_p$). The mass ratio can theoretically be constrained using LC data alone; however, we found that including the LC (which has vastly more data points than the RV) will result in degeneracy between $q$ and surface equipotential $\Omega$, ultimately failing to yield a good RV curve fit. 

As our RVs did not cover orbital phases near the eclipse, we did not observe the Rossiter–McLaughlin Effect \citep[RME:][]{1924ApJ....60...15R,1924ApJ....60...22M}, which normally allows for independent determination of the primary's rotational velocity and orbital inclination. Therefore, we assumed the rotational-to-orbital velocity ratio of the primary ($F_1$) and secondary ($F_2$) components to be fixed at $F_1=F_2 = 1$. This is despite a small difference between these velocities as indicated by $\langle v_p \sin i\rangle$ from LSD fitting (which becomes important later in Section \ref{sec: Residual Analysis}). This assumption does not affect the RV and light curve modeling. We set the inclination fixed at $90^\circ$ as an initial constraint, which is reasonable given the near-perfect v-shaped light curve. A circular orbit $(e=\omega_0\equiv0)$ is assumed throughout these analyses.

\begin{figure}
    \centering
    \includegraphics[width=\linewidth]{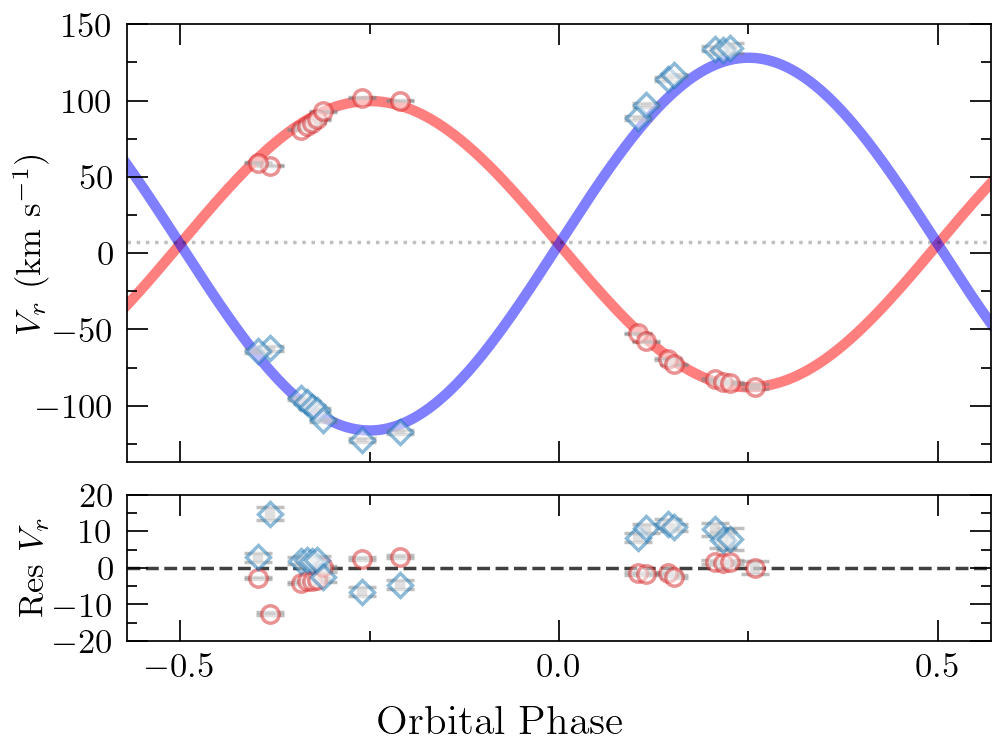}
    \caption{RV curves of V0757 Pup from MRES spectra. Red circles and blue squares represent the primary and secondary components, respectively, along with their modeled RVs from \texttt{PyWD2015} shown as solid red and blue curves. The dotted line denote the barycentric velocity of this system $(\gamma_\text{sys})$. The lower panel displays the residuals.}
    \label{fig:rvfit}
\end{figure}

After obtaining the best fit for $a,\;\gamma_\text{sys},\text{ and }q$, we locked them before proceeding to model the LCs. The resulting parameters are given in Table \ref{tab:param_output_LCRVphot}. The best fit for the RV curve is shown in Figure \ref{fig:rvfit}.

We then included all LCs and performed simultaneous RV and LC modeling. The primary effective temperature $T_1$ was fixed at $7033$ K (Section \ref{ssec:stellarparam}), while the secondary effective temperature, $T_2$, was set as an adjustable parameter. Based on the estimated effective temperatures, we assumed both stars have convective outer layers. Therefore, gravity darkening ($g$) and bolometric albedo ($A$) were fixed at canonical values: $g = 0.32$ \citep{Rucinski1969} and $A = 0.5$ \citep{Lucy1967}. We assumed a linear cosine limb-darkening law and allowed \texttt{PyWD2015} to automatically interpolate coefficients $(x_1, x_2, y_1, y_2)$ from available tables \citep{1993AJ_vanHamme}.

Since this is a detached binary, we used "MODE 2" for modeling, which was the only mode that satisfied the conditions and achieved convergence. In this mode, the dimensionless surface potentials of the primary and secondary $(\Omega_1, \Omega_2)$ were set as free parameters. Grid parameters were set to $N_\text{coarse} = 30$ and $N_\text{fine} = 60$, assuming zero third light contribution $(l_3\equiv0)$.

We fitted all adjustable parameters ($T_2,\;i,\;l_{1,\text{TESS}},\;\Omega_{1,2}$) using the differential corrections (DC) method until iterative corrections were smaller than their formal uncertainties. However, for some parameters, values fluctuated around specific points. Therefore, we extended the iterations until all parameters settled or true convergence was achieved. We noticed that due to slight differences in the light curves across sectors, simultaneous fitting tended to heavily favor sectors with the most datapoints (Sectors 61 and 88), resulting in poor fits for Sectors 7 and 34. To mitigate this, we re-ran the modeling individually using the RV data alongside just a single TESS sector at a time.

\begin{figure*}[h!]
    \centering
    \includegraphics[width=\linewidth]{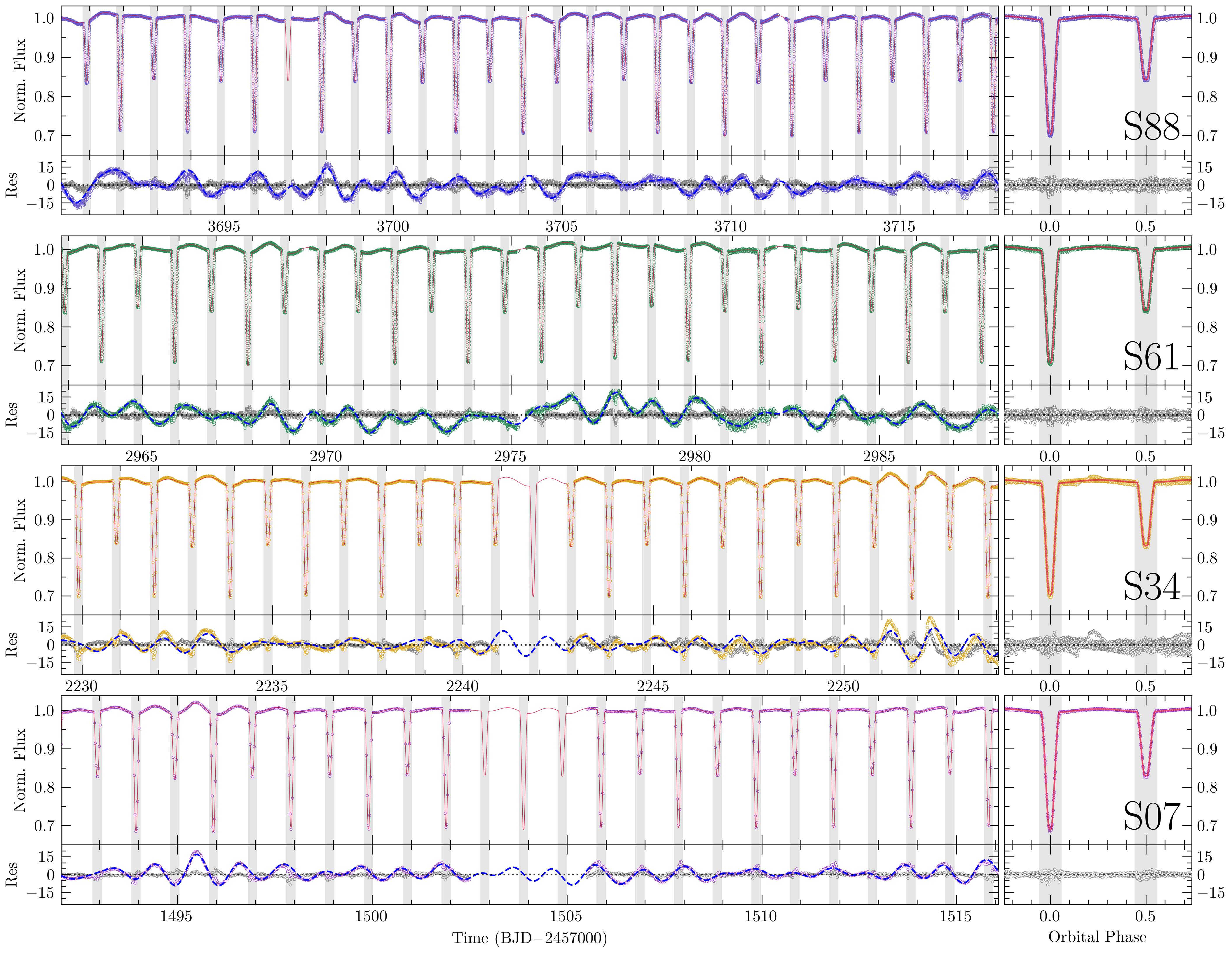}
    \caption{TESS light curves for each sector are shown as diffent coloured data points. \textbf{The left panels} are LC plots for each sector with X-axis is time in BTJD. The upper subplots shown combination of the best fit final \texttt{PyWD2015} model and the pulsational model (resulted from both group of frequencies) as a solid red curve. The lower subplots showed the residual light curve (after subtracting the eclipsing feature) shown as different colored data points and the "final" residual light curve (after subtracting both eclipsing and pulsational features) shown as gray data points. The blue dashed lines denoted the best fit pulsational light curve for each sectors. \textbf{The right panels} are phase-folded \textit{cleaned} light curve (after removing the pulsational signals) for each sectors on the upper subplots, while the lower subplots are the residuals. The shaded gray regions indicate orbital phase areas excluded from periodogram analysis. \textit{Alt text: Light curve fit of V0757 Pup for each sector}}
    \label{fig:lcfit}
    \vspace{-0.5cm}
\end{figure*}

The residual light curves showed fluctuations that were not caused by binarity, consistently appearing across all four sectors (see the left subplots of Figure \ref{fig:lcfit}). These fluctuations could be due to star spots or pulsations. Given the primary's high temperature and major flux contribution (>70\%), spot activity on the primary is unlikely. If spots were present on the secondary G1V component, fluctuations would exhibit periodic asymmetric flux in each quadrature. Therefore, we assumed that primary star pulsations are the main driver of these fluctuations, which may cause flux deflection during non-eclipse. 

Consequently, we performed residual analysis (prewhitening) to remove the pulsational signals (details in Section \ref{sec: Residual Analysis}). We synthesized a pulsational light curve from extracted frequencies and subtracted it from the original TESS LCs. The "cleaned" light curves were then reanalyzed using identical RV-LC modeling to obtain improved parameters. We noticed residual pulsation signals even in this "second" light curve, prompting a second prewhitening process (as signals were significant with S/N $\geq4$, except in Sector 34). We combined all extracted frequencies to improve the pulsation model, subtracted it from the original LCs, and produced a final \textit{relatively} clean light curve (see gray residual plots in Figure \ref{fig:lcfit}). Reanalyzing this final light curve yielded our final best-fit parameters for each sector. We then averaged $T_2,\;i,\;r_{1,2},\;\Omega_{1,2}$ across sectors for further analysis. Sector outputs and adopted parameters are shown in Table \ref{tab:param_output_LCRVphot}, and the final light curve fits are shown in Figure \ref{fig:lcfit}.

\begin{table*}[t!]
\centering
\caption{Output parameters of RV and LC curve modeling from \texttt{PyWD2015}}
\label{tab:param_output_LCRVphot}
\begin{tabular}{lccccc}
\hline
\textbf{Parameters} & \textbf{S07} & \textbf{S34} & \textbf{S61} & \textbf{S88} & \textbf{Adopted} \\[2pt] \hline
$T_0$$^\dagger$ (day)                      & \multicolumn{5}{c}{3717.7694446 (Fixed)}                                                  \\[2pt]
$P_\text{orb}$ (day)                       & \multicolumn{5}{c}{1.9891150 (Fixed)}                                                     \\[2pt]
$a$ ($R_\odot$)                            & \multicolumn{5}{c}{8.71(7)}                                                               \\[2pt]
$\gamma_\text{sys}$ ($\text{km s}^{-1}$)   & \multicolumn{5}{c}{5.57(75)}                                                              \\[2pt]
$q$ ($M_s/M_p$)                            & \multicolumn{5}{c}{0.716(1)}                                                              \\[2pt]
$i$ ($^\circ$)                             & 86.28(2)    & 86.36(3)    & 86.16(1)    & 86.15(1)    & 86.24(10)   \\[2pt]
$T_2$ (K)                                  & 5887(2)     & 5953(3)     & 5860(0)     & 5830(0)     & 5883(52)    \\[2pt]
$\Omega_1$                                 & 6.070(4)    & 6.057(5)    & 6.079(1)    & 6.079(2)    & 6.071(11)   \\[2pt]
$\Omega_2$                                 & 8.053(7)    & 7.761(9)    & 8.288(2)    & 8.266(2)    & 8.092(245)  \\[2pt]
$l_\text{1,TESS}$                          & 10.540(3)   & 10.550(4)   & 10.701(1)   & 10.699(1)   & --          \\[2pt]
$r_{1, \text{pole}}$                       & 0.18813(15) & 0.18667(17) & 0.18778(5)  & 0.18772(6)  & 0.1876(6)   \\[2pt]
$r_{1, \text{point}}$                      & 0.19100(16) & 0.18930(18) & 0.19064(6)  & 0.19057(6)  & 0.1904(7)   \\[2pt]
$r_{1, \text{side}}$                       & 0.18925(15) & 0.18773(18) & 0.18890(5)  & 0.18883(6)  & 0.1887(7)   \\[2pt]
$r_{1, \text{back}}$                       & 0.19059(16) & 0.18893(18) & 0.19023(5)  & 0.19016(6)  & 0.1900(7)   \\[2pt]
$r_{2, \text{pole}}$                       & 0.11056(11) & 0.10750(15) & 0.10692(4)  & 0.10722(4)  & 0.1081(17)  \\[2pt]
$r_{2, \text{point}}$                      & 0.11105(11) & 0.10796(15) & 0.10736(4)  & 0.10766(4)  & 0.1085(17)  \\[2pt]
$r_{2, \text{side}}$                       & 0.11073(11) & 0.10766(15) & 0.10708(4)  & 0.10738(4)  & 0.1082(17)  \\[2pt]
$r_{2, \text{back}}$                       & 0.11101(11) & 0.10792(15) & 0.10732(4)  & 0.10762(4)  & 0.1085(17)  \\[2pt] \hline
\multicolumn{6}{l}{$^{\dagger}$ $\text{BJD}-2.457\times10^6$} \\[2pt]
\end{tabular}
\end{table*}

\subsection{Absolute Parameters}\label{ssec:absolute_param}
Absolute parameters of the system are provided by \texttt{PyWD2015} but without uncertainty estimates. Therefore, we recalculated the absolute parameters using the results from Section \ref{ssec:rv_photo_fitting}. First, we calculated the semi-amplitude RV for each component $(K_1,\;K_2)$ using the values $a=a_1+a_2$, $P$, $q$, and $i$ derived previously. Assuming a circular orbit, we calculated these parameters and their propagated uncertainties via the relations:
\begin{align}
    a_1&=a\cdot\frac{q}{1+q}\\
    K_{1,2}&=\frac{2\pi\sin i}{P}\cdot a_{1,2}
\end{align}
We calculated all other absolute parameters ($M,\;R,\;\log g_\text{phot},\;L,\;M_{\text{bol}}$) with the \texttt{AbsParEB} software\footnote{\url{http://users.uoa.gr/~alliakos/Softwares/AbsParEB/}} \citep{2015ASPC_AbsPar_Liakos}: "Mode 1" (utilizing both spectroscopic and photometric data for double-lined spectroscopic binaries). Parameter errors were estimated through propagation of input parameter errors. The inputs provided to the program were the semi-amplitude RVs, effective temperatures (primary from Section \ref{ssec:stellarparam}, secondary from Table \ref{tab:param_output_LCRVphot}), fractional radii ($r_1,\;r_2$ from Table \ref{tab:param_output_LCRVphot}), orbital period, and orbital inclination. The derived parameters are tabulated in Table \ref{tab:abs_par}.

We note that the photometric surface gravity for the primary component is more precise and higher by $\sim0.13$ dex than the spectroscopic one, which is well within $\sim1\sigma$ of the $\log g_\text{spec}$ uncertainty. This difference may arise from imperfect normalization, especially around wide Balmer wings spanning $\sim100$\AA. Another method to derive $\log g$ spectroscopically is ionization balance analysis of Fe lines, which is beyond the scope of this study.

Next, we derived the photometric distance of V0757 Pup using the fundamental equation:
\begin{align}
\log d=0.2\mu+1=0.2\left(X-M_{X,\text{tot}}-A_X+5\right),\label{eq:fundamental}
\end{align}
where $X$ is the photometric band and $\mu$ is the distance modulus. We investigated the consistency between the Johnson-Cousins $V$ band and the \textit{Gaia} $G$ band. For the $V$-band, we used our derived reddening parameter (Section \ref{ssec:reddening}) and assumed the Galactic extinction law $A_V=3.1E(B-V)$ \citep{1989ApJ_cardelli}. For the $G$-band, we applied the relation $A_G=2.72E(B-V)$ \citep{2019AJ_stassun}.

We derived the total absolute magnitude for each band with these equations:
\begin{align}
    M_{V,j} &= M_{\text{bol},j}-BC_{V,j}\quad\quad j=\{p,s\}\\
    M_{V,\text{tot}} &=-2.5\log\left(10^{-0.4M_{V,p}}+10^{-0.4M_{V,s}}\right).
\end{align}
The bolometric correction (BC) for each component in each band was calculated through interpolation using bolometric correction grids provided by the \texttt{MIST} project via the \texttt{isochrones} package (see Section \ref{sec:evolution} for details). The BC interpolation employed temperature, surface gravity, metallicity, and bandpass extinction as inputs. For observed bandpass magnitudes, we adopted $V = 10.706 \pm 0.008$ mag \citep{2022yCat_TESSv82} and $G = 10.464 \pm 0.003$ mag \citep{2023A&A...674A...1GaiaDR3}. Uncertainties were estimated using Monte Carlo sampling with $10^6$ iterations. All derived parameters and uncertainties are presented in Table \ref{tab:abs_par}.

When comparing these photometric distances to the parallax distance, $d_\text{par} = 352.373 \pm 2.136$ pc \citep{2023A&A...674A...1GaiaDR3}, we observed offsets of $\sim36$ pc and 5 pc for the $V$ and $G$ bands, respectively. Both photometric distances agree reasonably well with the parallax distance and the distance derived from SED analysis (Section \ref{ssec:sedfitting}), within $1\sigma$ ($\sim$50 pc). Small discrepancies might arise because catalog apparent magnitudes may not reflect the actual maximum flux of both components, possibly due to observations near eclipse phases. Adjusting the magnitudes to be $\sim0.2$ mag brighter in the $V$ band would bring the distance perfectly in line with the \textit{Gaia} distance. Furthermore, BC interpolation may contribute uncertainty due to errors in atmospheric parameter inputs. Ultimately, we demonstrated an independent method for deriving distances based solely on radial velocity and light curve data, showing reasonable consistency with parallax distance.

\begin{table}[h!]
\centering
\caption{Derived Absolute parameters of V0757 Pup}
\label{tab:abs_par}
\begin{tabular}{lcr}
\hline
\multicolumn{1}{c}{\textbf{Parameters [unit]}} & \textbf{Primary}                            & \textbf{Secondary}                       \\[2pt] \hline
$K$ [km s$^{-1}$]                              & $92.204\pm0.791$                            & $128.809\pm1.104$                        \\[2pt]
                                               & \multicolumn{1}{l}{}                        & \multicolumn{1}{l}{}                     \\[2pt]
$M$ [$M_\odot$]                                & $1.305\pm0.026$                             & $0.934\pm0.030$                          \\[2pt]
$R$ [$R_\odot$]                                & $1.643\pm0.020$                             & $0.941\pm0.079$                          \\[2pt]
$\log g_\text{phot}$ [cm s$^{2}$]                                & $4.122\pm0.014$                             & $4.461\pm0.074$                          \\[2pt]
$L$ [$L_\odot$]                                & $5.918\pm0.269$                             & $0.951\pm0.166$                          \\[2pt]
$M_\text{bol}$ [mag]                           & $2.819\pm0.262$                             & $4.804\pm0.990$                          \\[2pt]
                                               & \multicolumn{1}{l}{}                        & \multicolumn{1}{l}{}                     \\[2pt]
A$_X$ [mag]                                    & \multicolumn{2}{c}{$0.079\pm0.012$ (V) ; $0.070\pm0.011$ (G)}       \\[2pt]
BC$_V$ [mag]                                   & $-0.066\pm0.014$                            & $-0.117\pm0.015$                         \\[2pt]
BC$_G$ [mag]                                   & $-0.023^{+0.012}_{-0.013}$                 & $-0.001\pm0.012$                         \\[2pt]
$M_\text{X,tot}$ [mag]                         & \multicolumn{2}{c}{$2.682^{+0.267}_{-0.283}$ (V) ; $2.628^{+0.269}_{-0.286}$ (G)} \\[2pt]
$\mu_\text{X}$ [mag]                           & \multicolumn{2}{c}{$7.944^{+0.284}_{-0.268}$ (V) ; $7.766^{+0.288}_{-0.269}$ (G)} \\[2pt]
$d_\text{phot}$ [pc]                           & \multicolumn{2}{c}{$387.94^{+54.21}_{-45.01}$ (V) ; $357.38^{+50.61}_{-41.58}$ (G)} \\[2pt]
\hline
\end{tabular}
\end{table}

\subsection{SED Fitting}\label{ssec:sedfitting}

In this section, we explore an alternative approach to estimating the system parameters of V0757 Pup through spectral energy distribution (SED) fitting. This method aims to reproduce observable parameters, such as stellar parameters and multiband photometric magnitudes. We employed \texttt{SPEEDYFIT}\footnote{\url{https://github.com/vosjo/speedyfit}} \citep{2017A&A...605A.109V}, which utilizes integrated SED flux from \citet{1979ApJS...40....1K} atmospheric models, alongside all available photometric data and the \textit{Gaia} DR3 parallax ($\varpi = 2.8610 \pm 0.0172$ mas) as inputs. To optimize parameters, this package employs Markov Chain Monte Carlo (MCMC) sampling.

During the analysis, we found it important to incorporate additional constraints, including the effective temperature from Section \ref{ssec:stellarparam} for the primary and from Table \ref{tab:param_output_LCRVphot} for the secondary, the mass ratio from Table \ref{tab:param_output_LCRVphot}, and the reddening parameters from Section \ref{ssec:reddening}. These refinements yielded results largely consistent with the light curve analysis.

\begin{figure*}[h!]
    \centering
    \includegraphics[width=\linewidth]{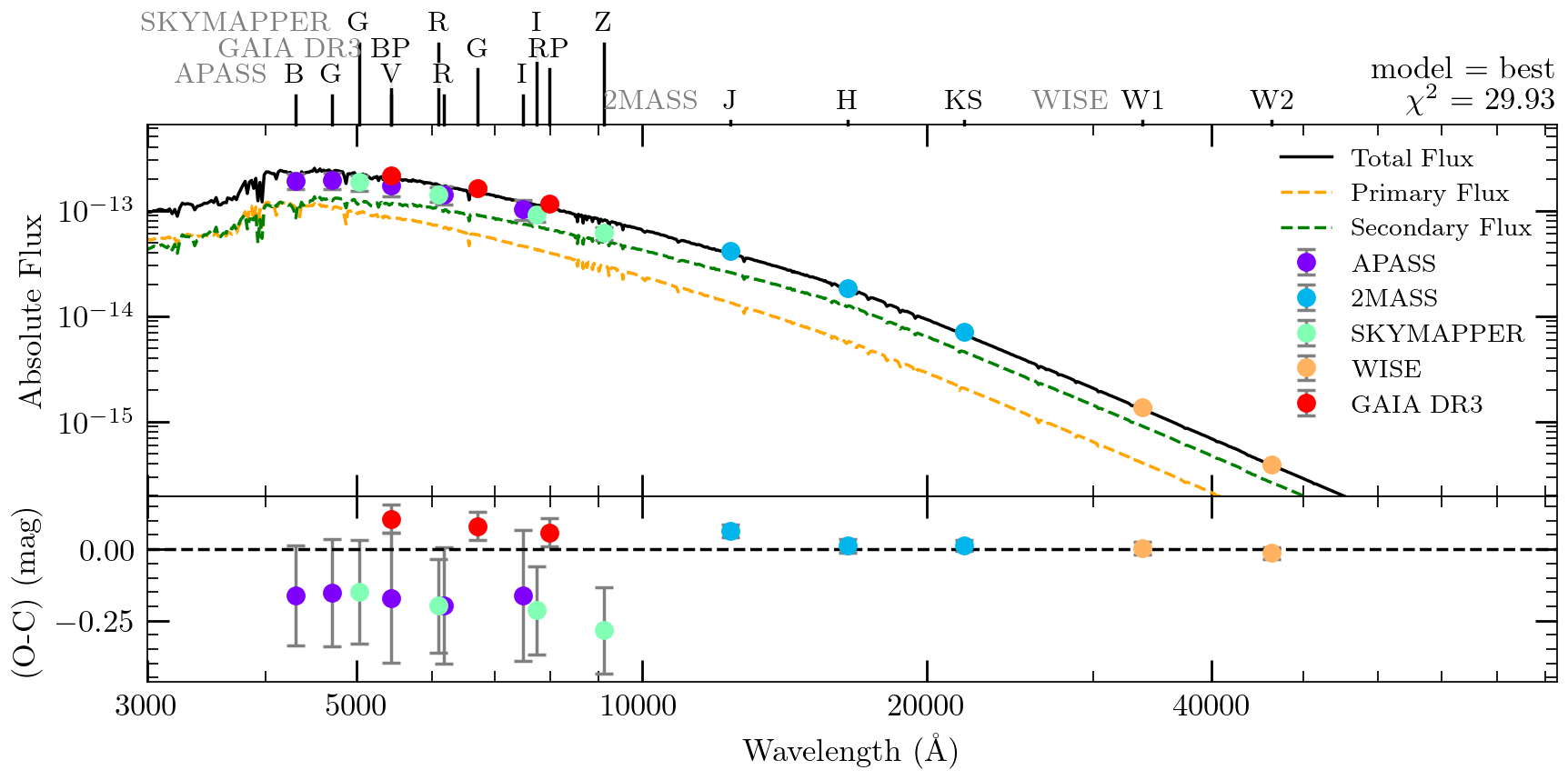}
    \caption{The Spectral Energy Distribution of V0757 Pup resulting from \texttt{SPEEDYFIT}. \textit{Alt text: SED graph of V0757 Pup}}
    \label{fig:sed_fitting_plot}
\end{figure*}

The SED fitting inputs and corresponding results are presented in Figure \ref{fig:sed_fitting_plot} and Table \ref{tab:speedyfit_result_input}. Our SED analysis estimates a distance of $d_\text{SED} = 349.59^{+2.13}_{-2.08}$ pc, slightly lower than the previous measurement of $d_\text{par} = 352.373\pm{2.136}$ pc \citep{2023A&A...674A...1GaiaDR3}, though both offer greater precision than the photometric distance (Section \ref{ssec:absolute_param}). Similarly, the derived stellar masses closely align with values obtained from radial velocity and light curve analysis. However, as shown in Figure \ref{fig:sed_fitting_plot}, most ground-optical band magnitudes are $\sim0.25$ mag lower than predicted, which significantly impacts derived temperatures and luminosities; the primary and secondary effective temperatures are each roughly 300 K lower.

Nevertheless, all derived values remain consistent within $2\sigma$, demonstrating the robustness of this approach. While this method provides a reliable initial estimate of stellar temperatures, photometric catalog magnitudes may not fully account for variability, such as eclipses or intrinsic stellar pulsations. The temperatures, stellar masses, and luminosities derived here will not be used in further analysis.

\begin{table}[h!]
\caption{SED Fitting observable inputs taken from \textit{Gaia DR3} \citep{2023A&A...674A...1GaiaDR3}, APASS \citep{2015AAS...22533616H}, SkyMapper \citep{2018PASA...35...10W}, 2MASS \citep{2006AJ....131.1163S}, and \textit{WISE} \citep{2010AJ_wise} and fitted results taken from median, 16th and 84th percentiles of the posterior distribution.}
\label{tab:speedyfit_result_input}
\centering
\begin{tabular}{lr}
\hline
\textbf{Observable} {[}mag{]} &  \\[2pt]
SKYMAPPER $G$ & $10.939\pm0.181$ \\[2pt]
SKYMAPPER $R$ & $10.798\pm0.165$ \\[2pt]
SKYMAPPER $I$ & $10.746\pm0.155$ \\[2pt]
SKYMAPPER $Z$ & $10.833\pm0.151$ \\[2pt]
\textit{Gaia} DR3 $G$ & $10.464\pm0.003$ \\[2pt]
\textit{Gaia} DR3 $BP$ & $10.681\pm0.004$ \\[2pt]
\textit{Gaia} DR3 $RP$ & $10.094\pm0.004$ \\[2pt]
APASS $B$ & $11.286\pm0.175$ \\[2pt]
APASS $V$ & $10.827\pm0.227$ \\[2pt]
APASS $G$ & $11.015\pm0.187$ \\[2pt]
APASS $R$ & $10.753\pm0.205$ \\[2pt]
APASS $I$ & $10.643\pm0.230$ \\[2pt]
2MASS $J$ & $9.664\pm0.023$ \\[2pt]
2MASS $H$ & $9.493\pm0.023$ \\[2pt]
2MASS $K_S$ & $9.456\pm0.019$ \\[2pt]
WISE $W_1$ & $9.444\pm0.023$ \\[2pt]
WISE $W_2$ & $9.467\pm0.020$ \\[2pt] \hline
\textbf{Fitted parameters} &  \\[2pt]
$T_\text{eff p,s}$ {[}K{]} & $6708^{+218}_{-160}$ ; $5643^{+451}_{-712}$ \\[2pt]
$M_\text{p,s}$ {[}$M_\odot${]} & $1.37^{+0.80}_{-0.65}$ ; $1.01^{+0.59}_{-0.48}$ \\[2pt]
$L_\text{p,s}$ {[}$L_\odot${]} & $4.76^{+0.68}_{-1.31}$ ; $0.98^{+1.25}_{-0.64}$ \\[2pt]
$d_\text{SED}$ {[}pc{]} & $349.59^{+2.13}_{-2.08}$ \\[2pt] \hline
\end{tabular}
\end{table}

\section{Residual Analysis}\label{sec: Residual Analysis}
\begin{figure}[h!]
    \centering
    \includegraphics[width=\linewidth]{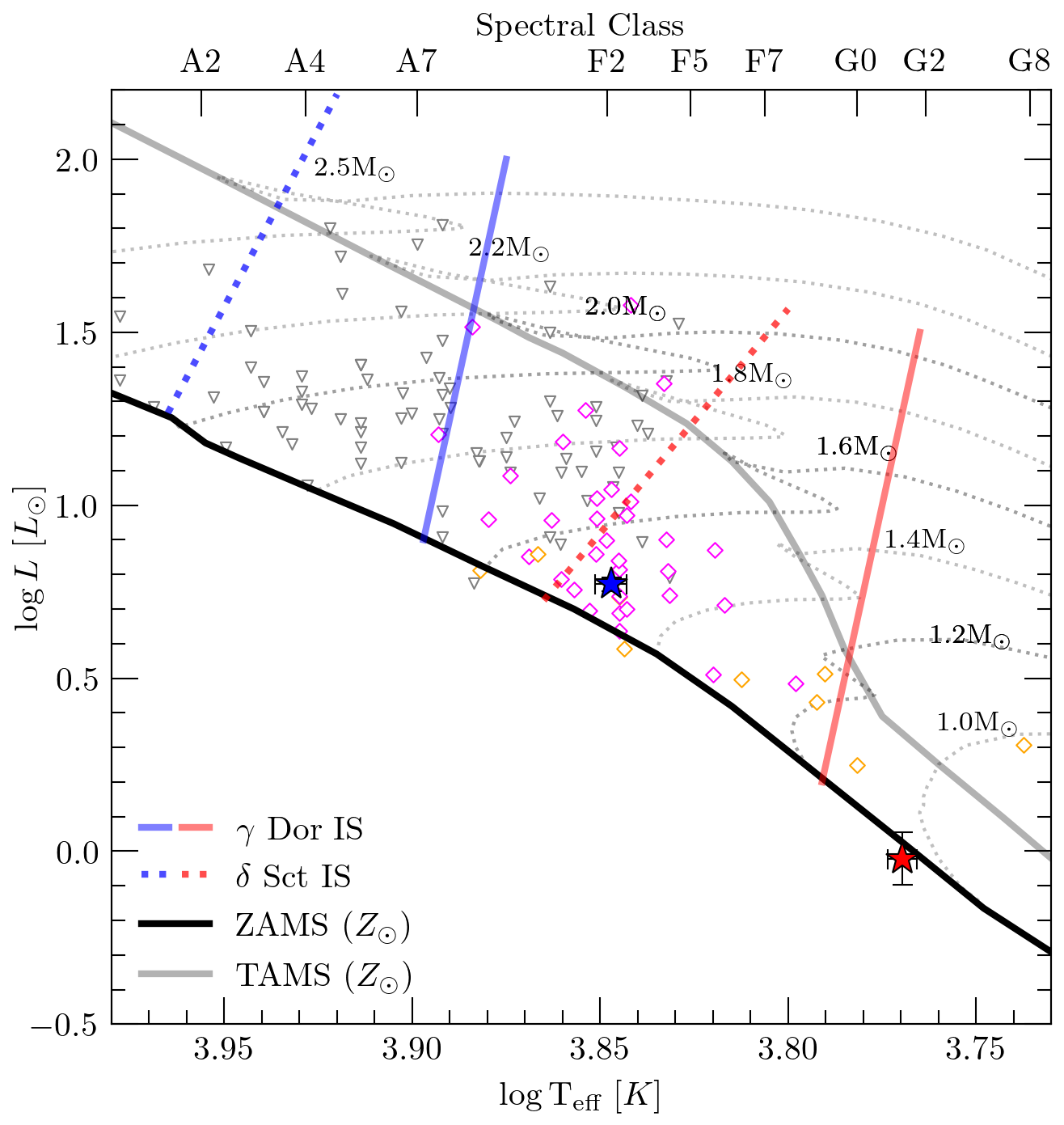}
    \caption{V0757 Pup's position on the HR diagram relative to similar systems. Black squares denote $\delta$ Sct stars in binaries \citep{liakos2016catalogue}, while magenta and orange diamonds represent the primary and secondary stars of total 44 $\gamma$ Dor in binaries (which orbital solution have been obtained), sourced from various studies \citep{2018NewA...62...70I,2018MNRAS.480.4693L,2018ApJ...865..115Z,2020MNRAS.491.5980H,2020RMxAA..56..321O,2025MNRAS.538..726Cakirli,2025PASJ..tmp..109L}. Solid black and grey lines indicate the theoretical zero-age main sequence (ZAMS) and terminal-age main sequence (TAMS, defined by core hydrogen exhaustion) for a solar metallicity single-star evolutionary model \citep{girardi2000evolutionary}. Dotted lines represent theoretical evolutionary tracks corresponding to each initial mass for solar metallicity, from the same references. Dashed blue and red lines mark the observational Instability Strip (IS) of $\delta$ Scuti \citep{murphy2019gaia}, while solid red and blue lines indicate the observational IS of $\gamma$ Doradus \citep{2025MNRAS.538..726Cakirli}. \textit{Alt text: Location of V0757 Pup primary and secondary component on Hertzsprung-Russel diagram}}
    \label{fig:HR_puls}
\end{figure}

As mentioned previously, residual analysis was performed iteratively to investigate intrinsic stellar variability—particularly stellar pulsations and other potential periodic features—ensuring all pulsational frequencies were extracted within detection limits. Based on the stellar parameters, we suggest the majority of pulsations originate from the primary F2V component, which resides within the $\gamma$ Dor instability strip (Figure \ref{fig:HR_puls}).

In general, multimode frequency analysis is performed on residual light curves to extract frequencies. For this, we used \texttt{MultiModes}, a script developed by \citet{Pamos2022}. This script performs mode fitting utilizing the \texttt{lmfit} minimizer \citep{2016ascl.soft06014N} to isolate the best frequency, performs a subtraction, and continues iteratively. We set a detection threshold of $\mathrm{SNR} \geq 4$ to confirm frequency reliability. We modified the SNR parameter calculation to better suit our needs, particularly in noise selection. The original script determines noise from a static minimum and maximum frequency range across all signals. We adapted this so noise is dependent on signal location by defining a $\mathrm{5 c/d}$ box around the detected frequency and applying 3$\sigma$ clipping to isolate noise.

\begin{figure*}[h!]
    \centering
    \includegraphics[width=\linewidth]{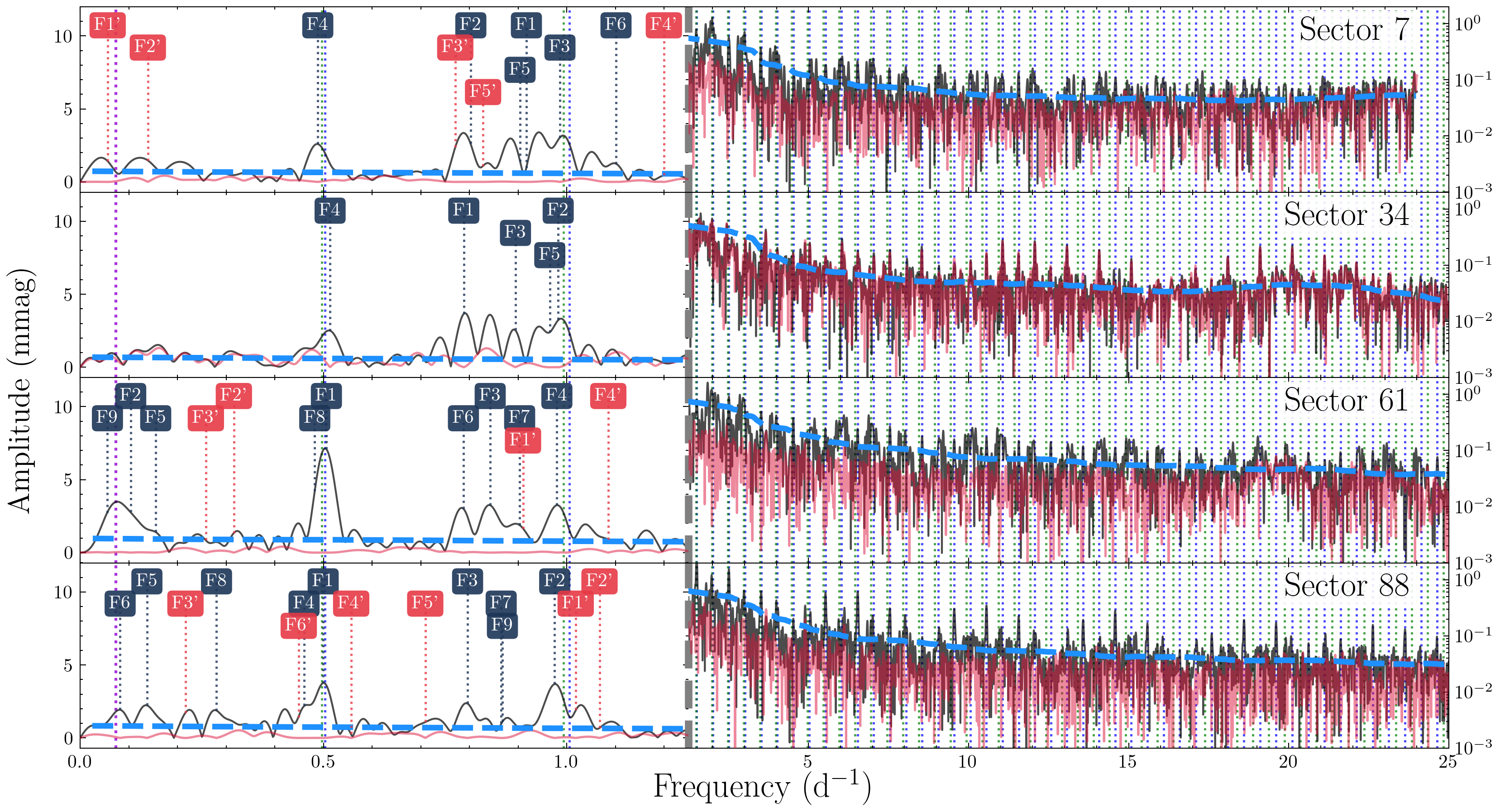}
    \caption{Periodograms for the residuals for each sector, before initial (black solid line) and after final (red solid line) dual prewhitening procesess with blue dashed line represent the noise level. For $f>1.25$ c/d, we change the Y-axis into log scale for better visualization. We annotated the location of detected first $(F_i)$ and second $(F_i')$ group frequencies as black and red vertical lines, respectively. We also indicated the location of $f_\text{orb}$ and $f_\text{rot}$ and their harmonics as blue and green dotted lines respectively. One peak near  $F_6$ on the Sector 88, is suspeciously close to the $f_\text{scat}\sim0.0737$ c/d (annotated as magenta dotted line).  \textit{Alt text: Periodogram from each sector (upper panel) and residual light curve fit of V0757 Pup that shows pulsation.}}
    \label{fig:periodogram}
\end{figure*}

We ran the residual analysis twice (after the first and second light curve modeling iterations) and masked the light curve around $\pm0.06$ in phase before and after the minima (see gray shades on Figure \ref{fig:lcfit}) to minimize the eclipsing features left over after LC modeling. These analyses yielded two sets of frequencies for Sectors 7, 61, and 88, denoted as $F_i$ and $F_i'$ in Table \ref{tab:detectedfrequency}. Sector 34 showed no significant signal above the S/N limit following the first LC modeling. The Fourier spectrum of each sector, both before the initial and after the final extraction, is shown in Figure \ref{fig:periodogram}. Pulsation signals synthesized using all detected frequencies for each sector are displayed in the left panels of Figure \ref{fig:lcfit}.

For multi-periodic pulsating stars, evaluating frequency combinations is particularly insightful, as they may arise due to the interaction of different independent oscillation modes. These combinations help explain nonlinear behavior and offer clues regarding internal physical processes \citep{SunXiaoYa2024} and evolutionary modeling \citep{Fitch1981, Lovekin&Guzik2017}. 

Initially, for the RV and LC modeling, we assumed synchronicity between the rotational and orbital periods of the primary. Here, we explored the possibility of asynchronicity by determining the stellar rotational frequency, $f_{rot}$, using the highly precise projected rotational velocity derived from our spectral analysis (Sec \ref{ssec:stellarparam}): $v_p\sin i=41.201\pm0.031$ km/s. The rotational frequency was calculated using the following relation:
\begin{align}
f_{rot}=\frac{\langle v_p\sin i\rangle}{2\pi R_\star\cdot\sin i}.
\end{align}
Incorporating the light curve data, we derived $f_{rot}=0.4967 \pm 0.0061$ c/d, which is extremely close—within $2\sigma$—to the orbital frequency. 

After identifying the frequencies, they were categorized as either independent, or combination of independent, orbital, or rotational frequencies. We employed the linear combination:
\begin{align}
    f_k=a\bullet f_i+b\bullet f_{j}+c\bullet f_{orb}+d\bullet f_{rot}\;,
\end{align}
with $[a,b,c,d]=\{-3,...,0,...,+3\}$ and the combination tolerance tailored per sector limited using 10\% of the Rayleigh criterion $(1/\Delta T)$ \citep{Guzik2022}. Orbital frequencies were prioritized, identifying them first before progressing to linear combinations. Our initial run suggested most frequencies could be explained as linear combinations, especially between $f_{orb}$ and $f_{rot}$. However, the large number of possible integer combinations makes accidental matches likely. Numerical agreement alone was considered insufficient evidence that a peak was a genuine combination frequency. Consequently, the resulting identifications are reported as "possible" combinations rather than definitive physical classifications.

\begin{figure*}[h!]
    \centering
\includegraphics[width=\linewidth]{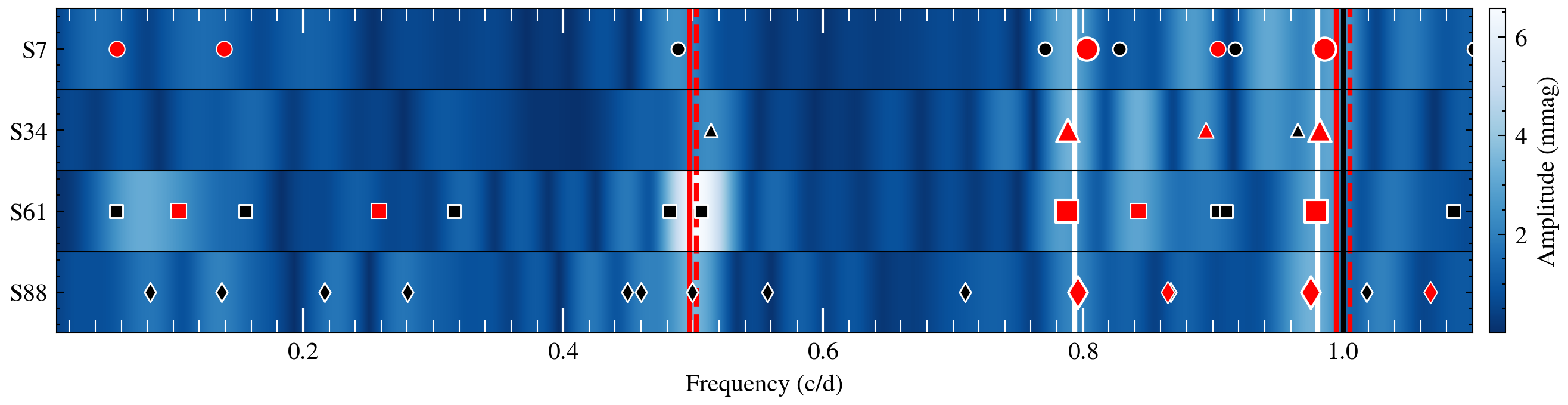}
    \caption{The periodogram of each sector is plotted as a heatmap, with the colour map representing amplitudes in mmag. Red-filled points indicate independent frequencies, while black-filled points represent combination frequencies. We have marked the orbital and rotational frequencies alongside their harmonics with dashed and solid red lines, respectively. Additionally, white solid-lines are added to indicate persistent independent frequencies while solid black lines indicate combination of rotation and orbital frequency. Black solid line indicate $f_\text{orb}+f_\text{rot}$. \textit{Alt text: Periodogram heatmap for four TESS sectors of V0757 Pup}}
    \label{fig:FrequencyCombination}
\end{figure*}

Figure~\ref{fig:FrequencyCombination} displays the initial Fourier spectrum for each sector for $f<1.5$ c/d, with signal amplitude as the intensity in mmag, alongside with the detected frequencies. Frequencies maintaining a consistent location across all four sectors were regarded as the strongest candidates for independent pulsation modes, while the remainder were classified as reference-related, possible combinations, or unclassified signals. Their measured values are listed in Table~\ref{tab:detectedfrequency}. The pulsation constants $Q$ for independent pulsation modes were then calculated via Equation \ref{eq:Q} \citep{Breger1990}, utilizing the parameters from Table \ref{tab:abs_par}:
\begin{equation}
\label{eq:Q}
    \log Q = \log P + \frac{1}{2} \log g + \frac{1}{10} M_\mathrm{bol} + \log T_\mathrm{eff} - 6.456
\end{equation}
which was derived from the relations $P\sqrt{\rho/\rho_{*}}=Q$ and $L\sim R^2 T_{eff}^4$.

Notably, low-frequency signals (periods around 10-20 days) were observed. These were neither classified as combination frequencies nor consistently present across all sectors (marked as "independent (?)" in Table \ref{tab:detectedfrequency}). A particular low-frequency signal near $\sim0.0737$ c/d ($f_\text{scat}$) in Sectors 61 and 88 might stem from scattering artifacts within the middle observation window (corresponding to $\sim13.7$ days). The absence of this signal in Sectors 7 and 34 likely results from our removal of two days of data in the middle of those sectors. Imperfections in the detrending algorithm during the reduction of Sectors 61 and 88 may have preserved this artifact. Nonetheless, some low-frequency signals were also observed as linear combinations of other frequencies.

The linear combinations in Table \ref{tab:detectedfrequency} represent only one of several possible configurations. However, based on our results, we confidently report that the most persistent frequencies across all sectors are $f\sim 0.79$ and $\sim0.98$ c/d, which we strongly suspect to be the fundamental pulsation frequencies (though the latter must be treated with caution due to its proximity to the possible $f_{orb}+f_{rot}\approx0.973$ linear combination). The calculated pulsational constants $Q$ were highly consistent, especially for high-significance frequencies, ranging from 0.5 to 0.6 days. This perfectly matches typical $\gamma$ Doradus $g$-mode pulsation values \citep{2010ApJ...713L.192G, 2011A&A...534A.125U}. We plotted the primary and secondary components on Hertzprung-Russell diagrams alongside 44 EA systems confirmed to have a $\gamma$ Doradus component, compiled from literature. The V0757 Pup primary sits squarely within the $\gamma$ Dor Instability Strip (IS) defined by \citet{2025MNRAS.538..726Cakirli}. While it also sits near the red edge of the observational $\delta$ Sct IS \citep{murphy2019gaia}, our analysis detected no higher frequencies ($>2$ c/d) with $S/N>4$ across any sector. Peaks on the initial Fourier spectrum are merely residuals of the binary signal, as evidenced by their coincidence with harmonics of $f_\text{orb}$ (see blue dotted lines in Figure \ref{fig:periodogram}).

\begin{table*}[h! ]
\centering
\caption{Detected original and combination pulsation frequencies per sector.}
\label{tab:detectedfrequency}
\begin{tabular}{c c c c c c c c}
\hline
ID & Frequency [c/d] & Amplitude [mmag] & Phase [rad] & SNR & Possible Combination & Q [d] \\[2pt]
\hline

\multicolumn{7}{c}{\textbf{Sector 7} ($\mathrm{tol}=0.0040956$, $f_{orb}=0.50273613$, $f_{rot}=0.4967$)} \\[2pt]
\hline

$F_{1}$ &
0.9170(1) &
5.89(4) &
0.059(1) &
9.5 &
$F_{5}+2f_{\rm orb}-2f_{\rm rot}$ &
\\[2pt]

$F_{2}$ &
0.8029(3) &
2.49(4) &
0.535(2) &
8.3 &
&
0.6738 \\[2pt]

$F_{3}$ &
0.9860(3) &
3.17(4) &
0.608(2) &
11.1 &
&
0.5486 \\[2pt]

$F_{4}$ &
0.4884(5) &
2.04(4) &
0.911(3) &
8.1 &
$-f_{\rm orb}+2f_{\rm rot}$ &
\\[2pt]

$F_{5}$ &
0.9040(1) &
6.53(4) &
0.195(1) &
12.4 &
independent (?) &
\\[2pt]

$F_{6}$ &
1.1012(8) &
1.19(4) &
0.550(5) &
8.7 &
$-F_{5}+3f_{\rm orb}+f_{\rm rot}$ &
\\[2pt]

$F_{1}^{\prime}$ &
0.0566(9) &
1.06(4) &
0.541(6) &
6.1 &
independent (?) &
\\[2pt]

$F_{2}^{\prime}$ &
0.1391(15) &
0.61(4) &
0.256(11) &
5.1 &
independent (?) &
\\[2pt]

$F_{3}^{\prime}$ &
0.7709(5) &
1.65(4) &
0.450(4) &
4.7 &
$2F_{2}^{\prime}+f_{\rm rot}$ &
\\[2pt]

$F_{4}^{\prime}$ &
1.1995(19) &
0.50(4) &
0.502(13) &
4.6 &
independent (?) &
\\[2pt]

$F_{5}^{\prime}$ &
0.8280(4) &
1.65(4) &
0.274(3) &
4.2 &
$-3F_{1}^{\prime}+f_{\rm orb}+f_{\rm rot}$ &
\\[2pt]

\hline

\multicolumn{7}{c}{\textbf{Sector 34} ($\mathrm{tol}=0.0039517$, $f_{orb}=0.50273613$, $f_{rot}=0.4967$)} \\[2pt]
\hline

$F_{1}$ &
0.7884(4) &
3.39(7) &
0.475(3) &
7.0 &
&
0.6861 \\[2pt]

$F_{2}$ &
0.9824(3) &
4.55(7) &
0.547(2) &
6.4 &
&
0.5506 \\[2pt]

$F_{3}$ &
0.8946(8) &
1.96(7) &
0.460(5) &
5.5 &
independent (?) &
\\[2pt]

$F_{4}$ &
0.5137(6) &
2.35(7) &
0.151(4) &
5.3 &
$-F_{2}+f_{\rm orb}+2f_{\rm rot}$ &
\\[2pt]

$F_{5}$ &
0.9651(4) &
3.77(7) &
0.709(3) &
5.5 &
$2F_{2}-f_{\rm orb}-f_{\rm rot}$ &
\\[2pt]

\hline

\multicolumn{7}{c}{\textbf{Sector 61} ($\mathrm{tol}=0.0039334$, $f_{orb}=0.50273613$, $f_{rot}=0.4967$)} \\[2pt]
\hline

$F_{1}$ &
0.5061(1) &
5.24(2) &
0.021(1) &
11.0 &
$f_{\rm orb}$ &
\\[2pt]

$F_{2}$ &
0.1040(1) &
2.77(2) &
0.653(1) &
6.1 &
independent (?) &
\\[2pt]

$F_{3}$ &
0.8426(1) &
3.27(2) &
0.970(1) &
8.2 &
independent (?) &
\\[2pt]

$F_{4}$ &
0.9796(1) &
3.74(2) &
0.061(1) &
9.9 &
&
0.5522 \\[2pt]

$F_{5}$ &
0.1553(2) &
1.97(2) &
0.751(2) &
8.1 &
$-F_{3}+f_{\rm orb}+f_{\rm rot}$ &
\\[2pt]

$F_{6}$ &
0.7877(2) &
2.09(2) &
0.892(1) &
8.3 &
&
0.6868 \\[2pt]

$F_{7}$ &
0.9032(1) &
6.24(2) &
0.181(0) &
5.7 &
$-F_{2}+2f_{\rm orb}$ &
\\[2pt]

$F_{8}$ &
0.4822(3) &
1.42(2) &
0.413(2) &
5.7 &
$F_{4}-f_{\rm rot}$ &
\\[2pt]

$F_{9}$ &
0.0562(1) &
3.57(2) &
0.927(1) &
5.3 &
$-F_{6}+F_{3}$ &
\\[2pt]

$F_{1}^{\prime}$ &
0.9101(1) &
6.45(2) &
0.085(0) &
5.7 &
$2F_{3}-3F_{3}^{\prime}$ &
\\[2pt]

$F_{2}^{\prime}$ &
0.3159(4) &
0.93(2) &
0.866(3) &
5.9 &
$3F_{2}+f_{\rm orb}-f_{\rm rot}$ &
\\[2pt]

$F_{3}^{\prime}$ &
0.2584(6) &
0.75(2) &
0.253(4) &
6.1 &
independent (?) &
\\[2pt]

$F_{4}^{\prime}$ &
1.0853(9) &
0.46(2) &
0.430(7) &
4.7 &
$F_{4}+F_{2}$ &
\\[2pt]

\hline

\multicolumn{7}{c}{\textbf{Sector 88} ($\mathrm{tol}=0.0036012$, $f_{orb}=0.50273613$, $f_{rot}=0.4967$)} \\[2pt]
\hline

$F_{1}$ &
0.4994(1) &
4.90(2) &
0.526(1) &
8.0 &
$f_{\rm rot}$ &
\\[2pt]

$F_{2}$ &
0.9752(1) &
3.70(2) &
0.092(1) &
10.9 &
&
0.5547 \\[2pt]

$F_{3}$ &
0.7962(2) &
2.34(2) &
0.037(1) &
8.5 &
&
0.6794 \\[2pt]

$F_{4}$ &
0.4602(1) &
4.72(2) &
0.603(1) &
6.3 &
$2F_{2}-3f_{\rm rot}$ &
\\[2pt]

$F_{5}$ &
0.1373(2) &
1.92(2) &
0.468(2) &
6.1 &
$-F_{7}+2f_{\rm orb}$ &
\\[2pt]

$F_{6}$ &
0.0822(2) &
1.79(2) &
0.522(2) &
6.0 &
$2F_{3}-3f_{\rm orb}\sim f_\text{scat}?$ &
\\[2pt]

$F_{7}$ &
0.8652(0) &
15.34(2) &
0.238(0) &
7.3 &
independent (?) &
\\[2pt]

$F_{8}$ &
0.2803(3) &
1.21(2) &
0.397(3) &
5.1 &
$F_{3}-3f_{\rm orb}+2f_{\rm rot}$ &
\\[2pt]

$F_{9}$ &
0.8678(0) &
15.50(2) &
0.385(0) &
5.4 &
$F_{7}$ &
\\[2pt]

$F_{1}^{\prime}$ &
1.0183(3) &
1.35(2) &
0.618(2) &
6.3 &
$-F_{2}+f_{\rm orb}+3f_{\rm rot}$ &
\\[2pt]

$F_{2}^{\prime}$ &
1.0673(4) &
1.07(2) &
0.335(3) &
6.4 &
independent (?) &
\\[2pt]

$F_{3}^{\prime}$ &
0.2165(4) &
0.93(2) &
0.036(3) &
5.7 &
$2F_{2}-2F_{7}$ &
\\[2pt]

$F_{4}^{\prime}$ &
0.5570(5) &
0.77(2) &
0.415(4) &
5.3 &
$F_{2}^{\prime}-2f_{\rm orb}+f_{\rm rot}$ &
\\[2pt]

$F_{5}^{\prime}$ &
0.7092(6) &
0.66(2) &
0.502(5) &
4.4 &
$-F_{3}+3f_{\rm orb}$ &
\\[2pt]

$F_{6}^{\prime}$ &
0.4492(1) &
3.54(2) &
0.498(1) &
4.1 &
$2F_{2}-2f_{\rm orb}-f_{\rm rot}$ &
\\[2pt]

\hline
\end{tabular}
\end{table*}

The photometric analysis results also confirmed that the secondary component is a G1V star. We suspected it might possess properties similar to those of the Sun, including solar pulsations. We attempted to detect the presence of this solar-type frequency pattern, known as solar-like oscillations \citep{SolarOscillation1, SolarOscillation2, SolarOscillation3}. However, after double prewhitening processes, we found no distinct comb-like pattern within an acceptable S/N. Therefore, we conclude that we cannot identify any pulsation source other than the primary $\gamma$ Doradus pulsations with the currently available data.


\section{Evolutionary Status}\label{sec:evolution}
Because V0757 Pup is a detached eclipsing binary, we assume that each component undergoes standard single-star evolutionary processes until one expands enough to fill its Roche Lobe. A comprehensive treatment of binary star evolution is beyond the scope of this study.

We began by interpolating evolutionary tracks using the \texttt{isochrones} package\footnote{\url{https://isochrones.readthedocs.io/en/latest/}} \citep{2015ascl.soft03010M}, based on the \textbf{M}ESA \textbf{I}sochrones and \textbf{S}tellar \textbf{T}racks \citep[\texttt{MIST}:][]{2016ApJS..222....8D,2016ApJ...823..102C}. This depends heavily on initial mass and metallicity estimates, which we took from our light curve and spectral analyses. Preliminary visual inspection revealed that the observed temperature, luminosity, and surface gravity were inconsistent with theoretical tracks given our uncertainties.

Because our mass measurements are highly reliable (supported by simultaneous RV and LC modeling), we suspect a systematic metallicity offset. To address this, we developed a routine to interpolate tracks using mass, metallicity, and evolutionary stage (represented as equivalent evolutionary phase, EEP) to minimize the difference between theoretical and observed $T_\text{eff}$, $\log g$, and $\log L$. We used the \texttt{emcee} sampler to optimize the log-likelihood, assuming Gaussian priors for $M$ (Table \ref{tab:abs_par}), $T_\text{eff}$ (primary from Table \ref{tab:best_fit_spectral}; secondary from Table \ref{tab:param_output_LCRVphot}), $\log g_\text{phot}$, and $\log L$ (Table \ref{tab:abs_par}), alongside a flat prior for metallicity.

The posterior distribution from this analysis is presented in Figure \ref{fig:mcmc_corner_evolution} in Appendix \ref{appendix:corner_plot}. Our results indicate that the LTE spectroscopic metallicity is $\sim0.26$ dex lower than the isochrone-derived metallicity. Referring to the 3D/NLTE study of Fe lines \citep{2022A&A...668A..68A}, for a star with atmospheric parameters closely matching V0757 Pup ($T_\text{eff} \sim 6500$ K, $\log g \sim 4.0$, $[\text{Fe/H}] \sim 0.0$ dex), the "true" 3D/NLTE Fe abundance is expected to be $\sim0.2$ dex higher than the 1D/LTE abundance. Note that different Fe lines with varying line strengths, excitation potentials, and ionization levels require different corrections. Therefore, considering the limitations of our spectroscopic analysis (lower spectral resolution, lack of line-by-line analysis, etc.), the difference between these metallicities could be entirely explained by non-LTE effects on Fe lines.


\begin{table}[h!]
\centering
\caption{Best fit parameters resulted from the isochrone analysis}
\label{tab:mcmc_params}
\begin{tabular}{lcc}
\hline
\multicolumn{1}{c}{\textbf{Parameters [unit]}} & \textbf{Primary}           & \textbf{Secondary}          \\[2pt] \hline
$\text{[Fe/H]}$ [dex]                          & $-0.367^{+0.055}_{-0.058}$ & $-0.244^{+0.071}_{-0.078}$  \\[2pt]
$M_\text{iso}$ [$M_\odot$]                     & $1.305\pm0.018$            & $0.940^{+0.020}_{-0.019}$   \\[2pt]
EEP                                            & $383.377^{+5.050}_{-5.668}$& $328.877^{+2.067}_{-2.029}$ \\[2pt]
                                               & \multicolumn{1}{l}{}       & \multicolumn{1}{l}{}        \\[2pt]
$\log$ Age [yr]                                & \multicolumn{2}{c}{$9.350^{+0.019}_{-0.021}$}             \\[2pt] \hline
\end{tabular}
\end{table}

\begin{figure*}[h!]
    \centering
    \includegraphics[width=\linewidth]{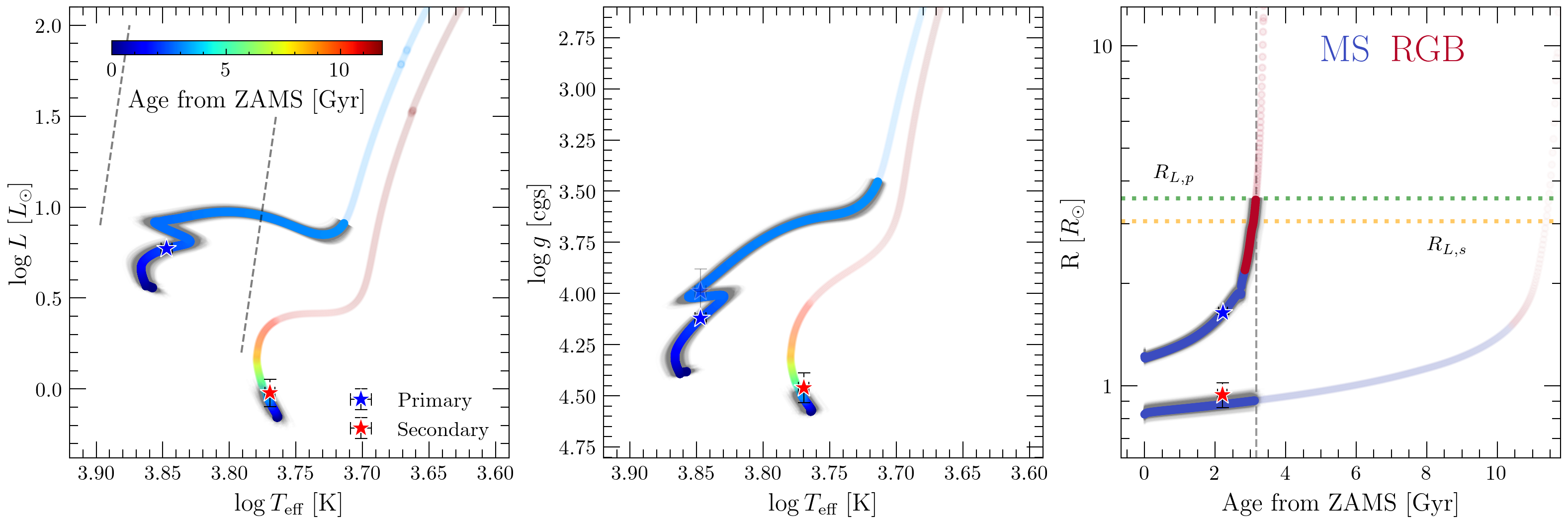}
    \caption{The \textbf{left} and \textbf{middle} panels display the \texttt{MESA} assumed single star evolutionary tracks of V0757 Pup on the Hertzsprung-Russell (HR) and Kiel diagrams, respectively. Dashed lines denote the boundary of $\gamma$-Dor IS \citep{2025MNRAS.538..726Cakirli}. The \textbf{right} panel illustrates the stellar radius evolution—primary on the left and secondary on the right—since reaching the Zero Age Main Sequence (ZAMS). The HR and Kiel tracks are colour-coded by age since ZAMS, while the radius evolution diagram uses colours to denote evolutionary stages: \textit{MS} (Main Sequence) and \textit{RGB} (Red Giant Branch). Shaded grey lines represent 1000 tracks accounting for mass and metallicity uncertainties, highlighting the track's variability. In the Kiel diagram, shaded points for $T_\text{eff}$ and $\log g_\text{spec}$ indicate that the photometric surface gravity remains within the $\sim1\sigma$ uncertainty of the spectroscopic measurement. Dotted green and orange horizontal lines are Roche lobe radius for primary and secondary component. Please note that these tracks will not be valid (indicated as shaded tracks) after mass-transfer event happen indicated as vertical dashed line on the right figure, see text for details. \textit{Alt text: Evolutionary track of V0757 Pup primary and secondary component on HR diagram (left), Kiel diagram (middle) and radius versus age (right) diagram.}}
    \label{fig:evol_literature}
\end{figure*}

Using the updated mass and metallicity, we generated evolutionary tracks for the primary and secondary components, shown in the Hertzsprung-Russell and Kiel diagrams in the left and middle panels of Figure \ref{fig:evol_literature}. These results place the system's estimated stellar age at $\sim2.2$ Gyr. The right panel of Figure \ref{fig:evol_literature} illustrates the radius evolution of each component since reaching the zero-age main sequence (ZAMS).

We also calculated the volume radius of the Roche lobe ($R_\text{L}$) for each component using equation \ref{eq: eggleton} from \cite{1983ApJ...268..368E}:
\begin{align}
\frac{R_\text{L,2}}{a}=\frac{0.49q^{2/3}}{0.6q^{2/3}+\ln(1+q^{1/3})},\label{eq: eggleton}
\end{align}
where $q \equiv M_2 / M_1$. The primary component will reach its Roche lobe first after evolving to the red giant branch (RGB) phase at a stellar age of about 3.04 Gyr, roughly 800 Myr from its current age. At this point, material from the primary will overflow to the secondary, initiating mass transfer event. V0757 Pup will then become a semi-detached binary system, and the evolutionary paths of both components will diverge drastically from the single-star tracks shown.

By this stage, the primary component will have moved well beyond the $\gamma$ Doradus instability strip (Fig. \ref{fig:HR_puls}) and will no longer exhibit pulsations. Comprehensive evolutionary modeling is required to explore this system's future, which will be undertaken in a future project.

\section{Conclusion}\label{sec:conclusion}

In this study, we have characterized V0757 Pup as a GDOR-EA system consisting of an F2V primary and a G1V secondary. By combining high-precision space-based photometry from four sectors of TESS observations with ground-based medium-resolution spectroscopy, we constrained the system's physical properties with high accuracy. Our spectroscopic analysis included the measurement of interstellar extinction using Na I D doublet absorption lines, yielding a reddening value of $E(B-V)=0.025^{+0.005}_{-0.004}$ mag, perfectly consistent with estimates from 2D-3D galactic extinction maps.

The orbital dynamics of the system were investigated using an O-C diagram spanning over 24 years, utilizing both archival minima and new ToM derived from TESS and TRT-net data. The O--C diagram revealed no significant deviations due to third body after correcting the linear trend. This conclusion is supported by Gaia DR3 astrometry, where the proper motion anomaly significance ($2.56\sigma$) fell below the $3\sigma$ detection threshold. 

Through simultaneous radial velocity and light curve modeling, precise absolute parameters has been derived for each component. The primary component of this system has a mass of $M_1=1.305\pm0.026M_\odot$ and a radius of $R_1=1.643\pm0.020R_\odot$, while the secondary is a solar analog with $M_2=0.934\pm0.030M_\odot$ and $R_2=0.941\pm0.079R_\odot$. We noted a systematic offset in metallicity estimates; our spectroscopic analysis indicates a sub-solar metallicity ([Fe/H]$_\text{spec}\approx-0.625$ dex), whereas isochrone fitting suggests a higher value by $\sim+0.2$ dex. While these values are consistent within $\sim1\sigma$ uncertainties, the difference could be explained by the 3D and non-LTE affecting Fe lines, demanding future high-resolution spectroscopic follow-up and more delicate chemical abundance analysis.

A key finding of this study is the confirmation of pulsation in the primary component. Analysis of the residual light curves, after removing the binary signature, revealed consistent independent frequencies at $f\approx0.79$ c/d and $0.98$ c/d across multiple TESS sectors. The derived pulsational constant ($Q\simeq0.5-0.6$ d) and the primary’s location within the theoretical instability strip on the Hertzsprung-Russell diagram firmly classify it as a $\gamma$ Doradus pulsator oscillating in \textit{g}-modes. Conversely, we found no evidence of solar-type oscillations in the G1V secondary component within the detection limits of the current data.

Finally, we utilized the derived physical parameters to determine the evolutionary status and distance of the system. The distances calculated from our orbital and SED modeling ($d\sim350$ pc) are in excellent agreement with the \textit{Gaia} DR3 parallax distance ($352.373\pm2.136$ pc), validating the robustness of our binary solution. Evolutionary tracks based on \texttt{MIST} models suggest a current age of $\sim2.2$ Gyr. The primary component is estimated to evolve off the main sequence and fill its Roche lobe in $\sim800$ Myr, at which point V0757 Pup will transition from a detached to a semi-detached binary system and the primary component is already left the instability strips.


Ultimately, V0757 Pup offers a unique laboratory for testing theoretical models of binary evolution and asteroseismology, promoting future high-precision photometric and spectroscopic campaigns of similar systems.

\section*{Supplementary Data}\label{supplement:ToM_data}
The following supplementary data is available at PASJ online.

E-table 2

\section*{Acknowledgments}
Based on observations made with MRES at the Thai National Observatory under program ID \texttt{TNTC012\_019} and the Thai Robotic Telescopes under program ID \texttt{TRTC11A\_009} and \texttt{TRTC11B\_010}, which are operated by the National Astronomical Research Institute of Thailand (Public Organization). DS and BR acknowledge and thank all related NARIT staff for their support and help during our observation and also during BR's visit to Doi Inthanon for onsite observation. BR also acknowledges the travel fund provided by the Leiden-Kerkhoven Bosscha Funds for the observation and thanks Feng Fabo and Xiao Guang-Yao for their helpful insights and discussions on the possibility of a third companion. This paper includes data collected by the TESS mission, obtained from the MAST data archive at the Space Telescope Science Institute (STScI). Funding for the TESS mission is provided by the NASA Explorer Program. STScI is operated by the Association of Universities for Research in Astronomy, Inc., under NASA contract NAS 5–26555. This research also made use of \texttt{Astropy}, a community-developed core Python package for Astronomy \citep{2018AJ....156..123A, 2013A&A...558A..33A}.

\section*{Funding}
We gratefully acknowledge the funding from the ITB research
grant under the PPMI 2024 Program (P.I: HLM).

\bibliography{ref.bib}
\bibliographystyle{apj.bst}

\vspace{-0.5cm}
\appendix 
\section{Radial Velocity \& Spectroscopic Observation Logs}\label{appendix:RV_table_output}

Table \ref{tab:RV} lists the individual radial velocity and projected rotational velocity measurements derived from the LSD profile fitting (see Section \ref{ssec:RVmeasurement}) for the primary and secondary components, respectively. The signal-to-noise of MRES spectra in each epoch, calculated around 600nm, are also included in this table.

\vspace{0.4cm}

\begin{table*}[t!]
\centering
\caption{The S/N of MRES spectra, radial velocity and projected rotational velocity of the V0757 Pup Primary and Secondary component measured from LSD profile fitting.}
\label{tab:RV}
\setlength{\tabcolsep}{4pt} 
\begin{tabular}{cccccccccccc}
    \hline
    BJD$-2457000$ & $\text{RV}_p$ & $\sigma(\text{RV}_p)$ & $v_p\sin i$ & $\sigma(v_p\sin i)$ & $\text{lf}_p$ & $\text{RV}_s$ & $\sigma(\text{RV}_s)$ & $v_s\sin i$ & $\sigma(v_s\sin i)$ & $\text{lf}_s$ & S/N \\[2pt]
    {[}d{]} & {[}km/s{]} & {[}km/s{]} & {[}km/s{]} & {[}km/s{]} & & {[}km/s{]} & {[}km/s{]} & {[}km/s{]} & {[}km/s{]} & & @600nm \\[2pt] \hline
3650.34648569 & -52.65 & 0.34 & 41.0 & 0.8 & 0.788 & 88.51 & 1.25 & 30.0 & 2.8 & 0.212 & 66 \\[2pt]
3650.36763275 & -57.59 & 0.32 & 40.9 & 0.7 & 0.797 & 97.39 & 1.26 & 30.0 & 2.8 & 0.203 & 60 \\[2pt]
3650.42657132 & -69.38 & 0.36 & 41.7 & 0.8 & 0.781 & 114.25 & 1.25 & 32.5 & 0.1 & 0.219 & 60 \\[2pt]
3650.44072722 & -72.84 & 0.31 & 45.0 & 0.6 & 0.785 & 116.54 & 1.00 & 29.9 & 2.5 & 0.215 & 59 \\[2pt]
3651.37091955 & 57.21 & 0.37 & 40.0 & 0.7 & 0.793 & -62.40 & 1.71 & 26.7 & 4.3 & 0.207 & 49 \\[2pt]
3651.45183845 & 80.81 & 0.27 & 40.0 & 0.0 & 0.776 & -95.14 & 0.89 & 30.0 & 2.3 & 0.224 & 51 \\[2pt]
3651.46602906 & 83.57 & 0.30 & 40.2 & 0.6 & 0.783 & -97.79 & 1.11 & 30.0 & 2.4 & 0.217 & 51 \\[2pt]
3651.48020809 & 85.59 & 0.34 & 40.0 & 0.7 & 0.793 & -100.67 & 1.17 & 34.1 & 2.4 & 0.207 & 60 \\[2pt]
3651.49439869 & 87.85 & 0.36 & 40.0 & 0.7 & 0.786 & -102.99 & 1.19 & 34.7 & 2.7 & 0.214 & 56 \\[2pt]
3693.07894428$^{\dag}$ & 25.84 & 0.39 & 95.0 & 0.7 & -- & -- & -- & -- & -- & -- & 74 \\[2pt]
3693.10861452 & 59.36 & 0.32 & 38.6 & 0.6 & 0.800 & -64.39 & 1.25 & 28.8 & 2.5 & 0.200 & 55 \\[2pt]
3693.27940170 & 92.91 & 0.38 & 37.0 & 0.9 & 0.786 & -109.63 & 1.26 & 30.0 & 2.8 & 0.214 & 41 \\[2pt]
3732.10301176 & -82.63 & 0.51 & 40.0 & 1.0 & 0.773 & 134.05 & 1.80 & 30.4 & 4.2 & 0.227 & 24 \\[2pt]
3732.12357781 & -84.57 & 0.51 & 35.3 & 1.1 & 0.785 & 133.03 & 2.17 & 26.1 & 5.6 & 0.215 & 30 \\[2pt]
3732.14121577 & -85.07 & 0.60 & 35.0 & 1.3 & 0.815 & 134.46 & 3.00 & 25.2 & 7.2 & 0.185 & 19 \\[2pt]
3732.20753169$^{\dag}$ & -87.63 & 1.71 & 48.8 & 2.9 & -- & 150.00 & 27.98 & 50.0 & 2.0 & -- & 7 \\[2pt]
3747.08961963 & 101.98 & 0.41 & 42.5 & 0.1 & 0.762 & -122.54 & 1.23 & 30.1 & 2.7 & 0.238 & 36 \\[2pt]
3747.18764537 & 99.90 & 0.32 & 40.0 & 0.0 & 0.752 & -117.11 & 1.21 & 25.0 & 2.8 & 0.248 & 36 \\[2pt] \hline
\textbf{Average} & -- & -- & \textbf{39.8} & \textbf{2.5} & \textbf{0.785} & -- & -- & \textbf{29.6} & \textbf{2.8} & \textbf{0.215} & -- \\[2pt] \hline
\multicolumn{12}{l}{\scriptsize $^{\dag}$ These epochs were excluded from spectral disentangling, RV modeling, and average calculations.} \\[2pt]
\end{tabular}
\end{table*}

\vspace{0.4cm}


\clearpage

\section{Corner Plots}\label{appendix:corner_plot}
\begin{figure*}[h!]
    \centering
    \includegraphics[width=\linewidth]{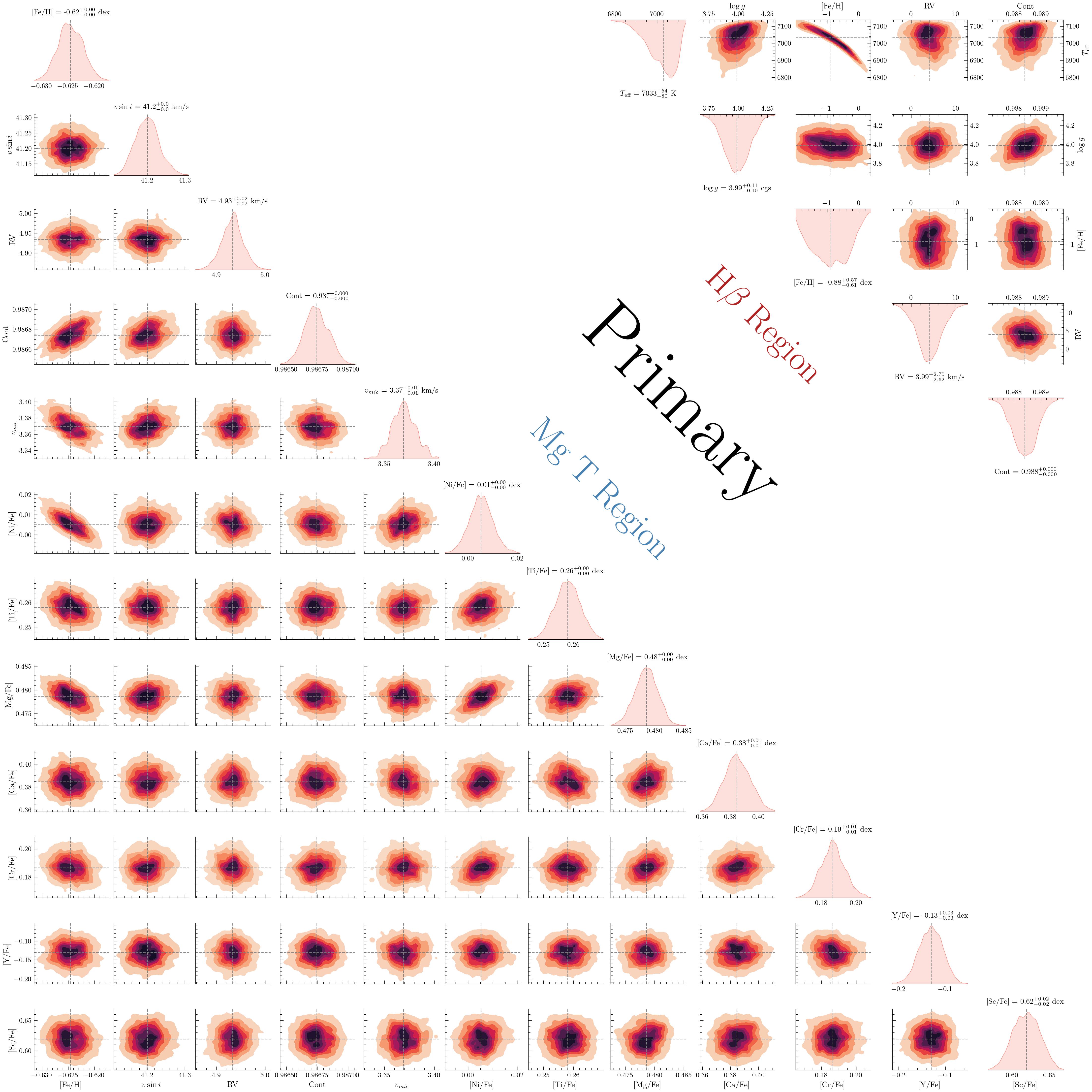}
    \caption{Posterior distributions from the spectral analysis of the V0757 Pup primary component's H$\beta$ region (upper right corner plots) and Mg triplet region (lower left corner plots) in V0757 Pup. Dashed crosshairs mark parameter medians; uncertainties represent the 16th and 84th percentiles. \textit{Alt text: Corner plot of spectral analysis on the primary spectra}}
    \label{fig:cornerbalmer}
\end{figure*}

\begin{figure}[h!]
    \centering
    \includegraphics[width=\linewidth]{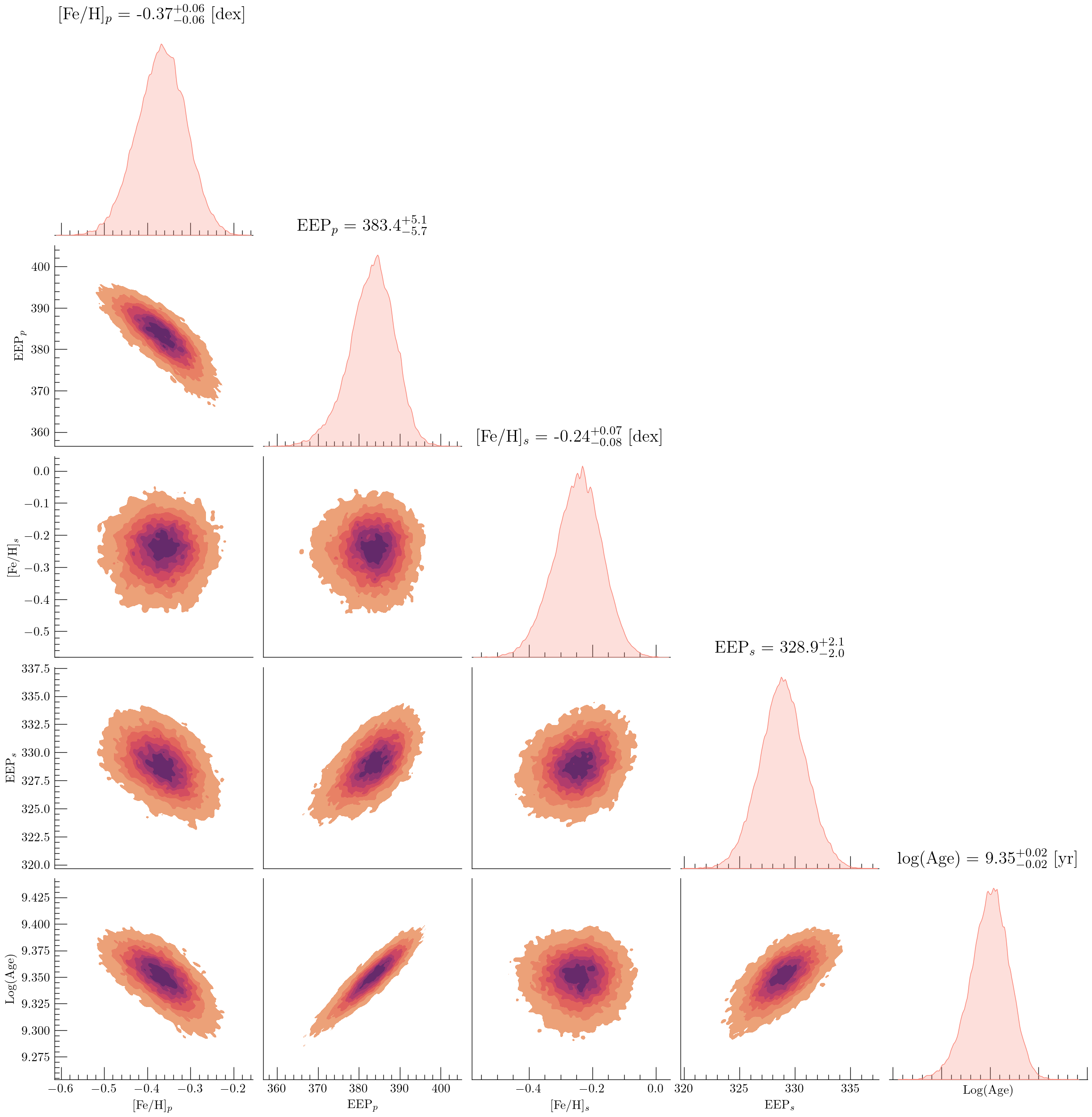}
    \caption{Posterior distribution of stellar evolutionary analysis with \texttt{isochrones} package. \textit{Alt text: Corner plot of stellar evolutionary analysis.}}
    \label{fig:mcmc_corner_evolution}
\end{figure}

We presented the corner plots of posterior distributions from MCMC analysis for the spectral analysis of primary component (see Section \ref{ssec:stellarparam}, shown in Figure \ref{fig:cornerbalmer}) and evolutionary analysis (see Section \ref{sec:evolution}, shown in Figure \ref{fig:mcmc_corner_evolution}).

\end{document}